\documentclass{article}
\usepackage{amsmath}
\usepackage{epsfig}

\numberwithin{equation} {section}

\begin{document}

\title{Proton spin structure and quark-parton model}
\author{Petr Z\'avada \\
\\
\textit{Institute of Physics, Academy of Sciences of the Czech Republic,}\\
\textit{Na Slovance 2, CZ-182 21 Prague 8}\\
email: zavada@fzu.cz}
\date{October 30, 1998}
\maketitle

\begin{abstract}
The alternative to the standard formulation of the quark-parton model is
proposed. Our relativistically covariant approach is based on the solution
of the master equations relating the structure and distribution functions,
which consistently takes into account the intrinsic quark motion. It is
suggested, that the intrinsic quark motion can substantially reduce the
structure function $g_1$. Simultaneously it is suggested, that the combined
analysis of the polarized and unpolarized data can give an information about
the effective masses and intrinsic motion of the quarks inside the nucleon.
\end{abstract}

\section{Introduction}

\label{sec1}

Measuring of the structure functions is an unique tool for the study of the
nucleon internal structure - together with the quark-parton model (QPM)
giving the relations between the structure functions and the parton
distributions, which represent a final, detailed picture of the nucleon. In
this sense, these relations, obtained under definite assumptions, are
extraordinary important, since the distribution functions themselves are not
directly measurable. At the same time, the standard, simple formulas
relating the structure and distribution functions are ordinarily considered
so self-evident, that in some statements, the both are identified.

The experiments dedicated to the Deep Inelastic Scattering (DIS), are
oriented to the measuring either \textit{unpolarized} or \textit{polarized}
structure functions. The results on the unpolarized functions are well
compatible with our expectations based on the QPM and QCD, but the situation
for the polarized functions is much more complicated. Until now, it is not
well understood, why the integral of the proton spin structure function $%
g_{1}$ is substantially less, than expected from very natural assumption,
that the nucleon spin is created by the valence quarks. Presently, there is
a strong tendency to explain the missing part of the nucleon spin as a
contribution of the gluons. It has been also suggested, that the quark
orbital momentum can play some role as well. Nevertheless, a consistent
explanation of the underlying mechanisms is still missing. During the last
years, the hundreds of papers have been devoted to the nucleon spin
structure, for the present status see e.g. \cite{jer},\cite{dis}, the
comprehensive overviews \cite{ans},\cite{hai} and citations therein.

In the present paper we summarize and update our discussion started in \cite%
{zav1}- \cite{zav3}, where we have shown, that the standard formulation of
the QPM, conceptually firmly connected with the infinite momentum frame
(IMF), oversimplifies the parton kinematics. In \cite{zav2} we demonstrated
that the effect of oversimplified kinematics in IMF can have an impact
particularly on the spin structure function $g_1$, or more exactly, it can
substantially modify the relation between the distribution and spin
structure functions.

The paper is organized as follows. In the following section the basic
kinematical quantities related to the DIS are introduced and particularly
the meaning of variable $x_B$ is discussed. In the Sec. \ref{seca3} we
consider proton as an idealized system of the quasifree, massive partons
with the four-momenta on the mass shell. In the covariant formulation we
deduce the relations between the structure and distribution functions for
the unpolarized and polarized case. At the same time, the obtained relations
are compared with those derived in the standard, IMF approach. In Sec. \ref%
{sec3}., using the results obtained in the previous sections, we propose a
more realistic model of the proton, in which the internal motion of quarks
is consistently taken into account. In contradistinction to the standard
treatment based on the QCD evolution of the distribution functions in
dependence on $Q^2$, our model rather aims at describing the part of
distribution and structure function, which is not calculable in terms of the
pertubative QCD. In the Sec. \ref{sec4}. the results of the model on the
polarized and unpolarized proton structure functions are compared with the
experimental data and some free parameters are fixed. Some additional
comments on the model and obtained results are done in the Sec. \ref{sec5}.
The last section is devoted to the summary and concluding remarks.

\section{Kinematics}

\begin{figure}[tbp]
\begin{center}
\epsfig{file=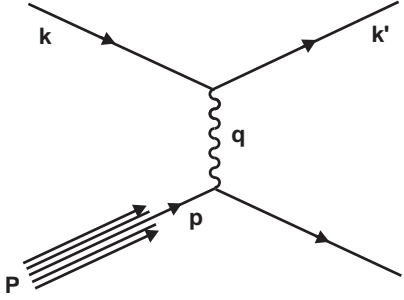, height=4cm}
\end{center}
\caption{Diagram describing DIS as a one photon exchange between the charged
lepton and parton.}
\label{yy1}
\end{figure}

First of all let us recall some basic notions used in the description of DIS
and the interpretation of the experimental data on the basis of the QPM. The
process is usually described (see Fig. \ref{yy1}) by the variables 
\begin{equation}  \label{k1}
q^2\equiv -Q^2=(k-k^{\prime })^2,\qquad x_B=\frac{Q^2}{2Pq}.
\end{equation}
As a rule, lepton mass is neglected, i.e. $k^2=k^{\prime 2}=0$. Important
assumption of the QPM is that the struck parton remains on-shell, that
implies 
\begin{equation}  \label{k2}
q^2+2pq=0.
\end{equation}
Bjorken scaling variable $x_B$ can be interpreted as the fraction of the
nucleon momentum carried by the parton in the nucleon infinite momentum
frame (IMF). The motivation of this statement can be explained as follows.
Let us denote 
\begin{equation}  \label{k3}
p(lab)\equiv (p_0,p_1,p_2,p_3),\qquad P(lab)\equiv (M,0,0,0),\qquad
q(lab)\equiv (q_0,q_1,q_2,q_3)
\end{equation}
the momenta of the parton, nucleon and exchanged photon in the nucleon rest
frame (LAB). The Lorentz boost to the IMF (in the direction of collision
axis) gives 
\begin{equation}  \label{k4}
p(\inf )\equiv (p_0^{\prime },p_1^{\prime },p_2,p_3),\qquad P(\inf )\equiv
(P_0^{\prime },P_1^{\prime },0,0),\qquad q(\inf )\equiv (q_0^{\prime
},q_1^{\prime },q_2,q_3),
\end{equation}
where for $\beta \rightarrow -1$ 
\begin{equation}  \label{k5}
p_0^{\prime }=p_1^{\prime }=\gamma (p_0+p_1),\quad P_0^{\prime }=P_1^{\prime
}=\gamma M,\qquad \gamma =1/\sqrt{1-\beta ^2}.
\end{equation}
If we denote 
\begin{equation}  \label{k6}
x\equiv \frac{p_0^{\prime }}{P_0^{\prime }}=\frac{p_1^{\prime }}{P_1^{\prime
}}=\frac{p_0+p_1}M,
\end{equation}
then one can write 
\begin{equation}  \label{k7}
p(\inf )=xP(\inf )+(0,0,p_2,p_3).
\end{equation}
Now let the lepton has initial momentum $k(lab)\equiv (k_0,-k_0,0,0)$. If we
denote $\nu \equiv k_0-k_0^{\prime }$ and $q_L\equiv q_1,$ then $q_L<0$ and
from Eqs. (\ref{k1}), (\ref{k2}) it follows 
\begin{equation}  \label{k8}
x_B=\frac{pq}{Pq}=\frac{p_0\nu +\mid q_L\mid p_1}{M\nu }-\frac{{\vec p}_T{%
\vec q}_T}{M\nu },
\end{equation}
where ${\vec p}_T,{\vec q}_T$ are the parton and photon transversal momenta.
Obviously 
\begin{equation*}
k^{\prime 2}=(k-q)^2=k^2+q^2-2k_0\nu +2k_0\mid q_L\mid =0,
\end{equation*}
\begin{equation}  \label{ka10}
\frac{\mid q_L\mid }\nu =1+\frac{Q^2}{2k_0\nu }=1+\frac M{k_0}x_B.
\end{equation}
Using this relation the Eq. (\ref{k8}) can be modified 
\begin{equation}  \label{ka8}
x_B=\frac{p_0+p_1}M+\frac{p_1}{k_0}x_B-\frac{{\vec p}_T{\vec q}_T }{M\nu }
\end{equation}
therefore, if the lepton energy is sufficiently high, so $p_1/k_0\approx 0$,
one can write 
\begin{equation}  \label{k9}
x_B=x-\frac{p_T\,q_T}{M\nu }\cos \varphi ,
\end{equation}
where $\varphi $ is the angle between the parton and photon momenta in the
transversal plane.

So, if parton transversal momenta are neglected, $x_{B}$ really represents
fraction of momentum (\ref{k6}) and (\ref{k7}). In a higher approximation
the experimentally measured $x_{B}$, being an integral over $\varphi $, is
effectively smeared with respect to the fraction $x-$ which is not
correlated with $\varphi .$ An estimation of the second term in the last
equation can be done as follows. Because 
\begin{equation*}
q^{2}=\nu ^{2}-\left| \vec{q}\right| ^{2},
\end{equation*}%
\begin{equation}
\left( \frac{\vec{q}}{\nu }\right) ^{2}=1+\frac{Q^{2}}{\nu ^{2}}=1+\frac{%
4M^{2}}{Q^{2}}x_{B}^{2},  \label{kb10}
\end{equation}%
then Eqs. (\ref{ka10}), (\ref{kb10}) give 
\begin{equation}
\frac{q_{T}}{\nu }=\sqrt{\left( \frac{\vec{q}}{\nu }\right) ^{2}-\left( 
\frac{q_{L}}{\nu }\right) ^{2}}=\sqrt{\left( \frac{4M^{2}}{Q^{2}}-\frac{M^{2}%
}{k_{0}^{2}}\right) x_{B}^{2}-\frac{2M}{k_{0}}x_{B}}<\frac{2Mx_{B}}{\sqrt{%
Q^{2}}};  \label{kba10}
\end{equation}%
therefore, for $M/k_{0}\approx 0$ we obtain 
\begin{equation}
\frac{pq}{M\nu }=\frac{p_{0}+p_{1}}{M}-\frac{2p_{T}x_{B}}{\sqrt{Q^{2}}}\cos
\varphi  \label{kd10}
\end{equation}%
and 
\begin{equation}
x_{B}=x-\frac{2p_{T}x_{B}}{\sqrt{Q^{2}}}\cos \varphi ,  \label{kc10}
\end{equation}%
which gives 
\begin{equation}
\frac{x}{x_{B}}=1+\frac{2p_{T}}{\sqrt{Q^{2}}}\cos \varphi .  \label{kci10}
\end{equation}%
i.e. $x=x_{B}$ for $Q^{2}\rightarrow \infty $. Therefore $x_{B}$ can be at
sufficiently high $Q^{2}$ considered as a good approximation of $x$ (and
vice versa). Accordingly, in the next sections we shall frequently use the
approximation 
\begin{equation}
\frac{pq}{M\nu }\approx \frac{p_{0}+p_{1}}{M},  \label{kcj10}
\end{equation}%
which will considerably simplify calculation of some integrals. The effect
of this approximation is well under control, all relevant relations could be
calculated without it, but at a price, that the sense of some obtained
expressions would be quite apparent only after a numeric calculation.

Now, if we assume parton phase space is spherical (in LAB) and an idealized
scenario in which the parton has a mass $m^2=p_0^2-p_1^2-p_2^2-p_3^2$, then
further relations can be obtained.

1) \textit{variable x}

\noindent From Eq. (\ref{k6}) and the condition $x\leq 1$ it can be shown 
\begin{equation}  \label{k11}
x\geq \frac{m^2}{M^2},
\end{equation}
\begin{equation}  \label{k12}
\sqrt{p_1^2+p_2^2+p_3^2}\leq p_m\equiv \frac{M^2-m^2}{2M},\qquad p_T^2\leq
M^2(x-\frac{m^2}{M^2})(1-x).
\end{equation}
Obviously, the highest value of $p_1$ is reached if $p_T=0$ and 
\begin{equation}  \label{ka12}
x=\frac{\sqrt{p_1^2+m^2}+p_1}M=1
\end{equation}
which gives 
\begin{equation}  \label{kb12}
p_{1\max }=p_m\equiv \frac{M^2-m^2}{2M}.
\end{equation}
Then spherical symmetry implies 
\begin{equation}  \label{kc12}
\sqrt{p_1^2+p_2^2+p_3^2}\leq p_m,
\end{equation}
i.e. the first relation in (\ref{k12}) is proved. Apparently, the minimal
value of $x$ is reached for $p_1=-p_m$ and $p_T=0$. After inserting to (\ref%
{k6}) one gets (\ref{k11}). Finally, the relation (\ref{k6}) implies 
\begin{equation}  \label{kd12}
p_1=\frac{M^2x^2-m^2-p_T^2}{2Mx}
\end{equation}
which, inserted to modified relation (\ref{kc12}) 
\begin{equation}  \label{ke12}
p_1^2+p_T^2\leq \left( \frac{M^2-m^2}{2M}\right) ^2
\end{equation}
after some computation gives the second relation in (\ref{k12}).

2) \textit{variable} $x_B$

\noindent Let us express $x_B$ in the LAB 
\begin{equation}  \label{k13}
x_B=\frac{pq}{Pq}=\frac{p_0\nu -\vec p\ \vec q}{M\nu }=\frac 1M\left( \sqrt{%
m^2+\left| \vec p\right| ^2}-\frac{\vec p\ \vec q}\nu \right)
\end{equation}
and estimate its minimal value. With the use of (\ref{kb10}) we obtain 
\begin{equation}  \label{k15}
x_B\geq \frac 1M\left( \sqrt{m^2+p_m^2}-p_m\sqrt{1+\frac{4M^2}{Q^2}x_B^2}%
\right) .
\end{equation}
Since 
\begin{equation}  \label{ka15}
\sqrt{1+\frac{4M^2}{Q^2}x_B^2}\leq 1+\frac{2M^2}{Q^2}x_B^2
\end{equation}
and 
\begin{equation}  \label{k16}
\frac 1M\left( \sqrt{m^2+p_m^2}-p_m\right) =\frac{m^2}{M^2},
\end{equation}
relation (\ref{k15}) can be rewritten 
\begin{equation}  \label{k17}
x_B\geq \frac{m^2}{M^2}-\frac{2Mp_m}{Q^2}x_B^2\geq \frac{m^2}{M^2}-\frac{%
2Mp_m}{Q^2}\frac{m^4}{M^4}=\frac{m^2}{M^2}\left( 1-\frac{2p_m}M\frac{m^2}{Q^2%
}\right) .
\end{equation}
i.e. for $m^2\ll Q^2$ lower limit of $x_B$ coincides with the limit (\ref%
{k11}).

\section{Idealized scenario: quasifree partons on mass shell}

\label{seca3}

In this section we imagine the partons as a gas (or a mixture of gases) of
quasifree particles filling up the nucleon volume. The prefix \textit{quasi }
here means that the partons bounded inside the nucleon behave at the
interaction with the external photon probing the nucleon as free particles
having the momenta on mass shell.

\subsection{Deconvolution of the distribution function}

Let us suppose $F(x)$ is the distribution function of some sort of partons
given in terms of variable $x$ according to Eq. (\ref{k6}) and these partons
are assumed to have the mass $m$. If the spherical symmetry is assumed in
the nucleon rest system and $G(p_0)d^3p$ is the number of partons in the
element of the phase space, then the distribution function $F(x)$ can be
expressed as the convolution 
\begin{equation}  \label{eq4}
F(x)=\int \delta \left( \frac{p_0+p_1}M-x\right) G(p_0)d^3p,\qquad p_0=\sqrt{%
m^2+p_1^2+p_2^2+p_3^2}.
\end{equation}
Using the set of integral variables $h,p_0,\varphi $ instead of $p_1,p_2,p_3$
\begin{equation}  \label{eq5}
p_1=h,\qquad p_2=\sqrt{p_0^2-m^2-h^2}\sin \varphi ,\qquad p_3= \sqrt{%
p_0^2-m^2-h^2}\cos \varphi ,
\end{equation}
the integral (\ref{eq4}) can rewritten 
\begin{equation}  \label{eq6}
F(x)=2\pi \int_m^{E_{\max }}\int_{-H}^{+H}\delta \left( \frac{p_0+h}%
M-x\right) G(p_0)p_0dhdp_0,\qquad H=\sqrt{p_0^2-m^2}.
\end{equation}
First, let us calculate inner integral within limits $\pm H$ depending on $%
p_0.$ For given $x$ and $p_0$ there contributes only $h$ for which 
\begin{equation}  \label{eq7}
p_0+h=Mx,
\end{equation}
but simultaneously $h$ must be inside the limits 
\begin{equation}  \label{eq8}
-\sqrt{p_0^2-m^2}\leq h\leq \sqrt{p_0^2-m^2}
\end{equation}
which means, that for 
\begin{equation}  \label{eq9}
p_0+\sqrt{p_0^2-m^2}<Mx
\end{equation}
or equivalently for 
\begin{equation}  \label{eq10}
p_0<\xi \equiv \frac{Mx}2+\frac{m^2}{2Mx}
\end{equation}
considered integral gives zero. For $p_0>\xi $, when the both conditions (%
\ref{eq7}), (\ref{eq8}) are compatible for some value $h$, the integral can
be evaluated 
\begin{equation}  \label{eq11}
\int_{-H}^{+H}\delta \left( \frac{p_0+h}M-x\right) G(p_0)p_0dh=MG(p_0)p_0.
\end{equation}
Therefore the integral (\ref{eq6}) can be expressed 
\begin{equation}  \label{eq12}
F(x)=2\pi M\int_\xi ^{E_{\max }}G(p_0)p_0dp_0.
\end{equation}
Let us note, the equation similar to this appears already in \cite{fra1} but
with the structure function $F_2(x)$ instead of the distribution one. We
shall deal with the $F_2$ in the next section, where it will be shown, that
the corresponding relation is more complicated. For a comparison see also %
\cite{cle}, where on the place of $G(p_0)$ the statistical distribution
characterized by some temperature and chemical potential is used.

\begin{figure}[tbp]
\begin{center}
\epsfig{file=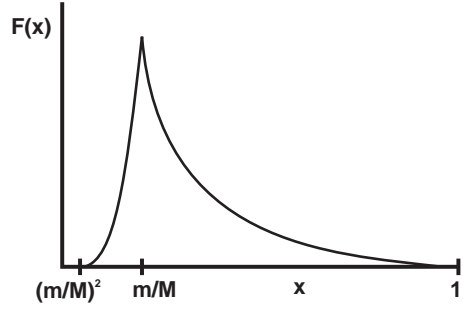, height=5cm}
\end{center}
\caption{Example of the function obeying Eqs. (\ref{eq24}), (\ref{eq24a}).}
\label{yy2}
\end{figure}

Next, from the relation (\ref{eq10}) we can express $x$ as a function $\xi $ 
\begin{equation}  \label{eq13}
x_{\pm }=\frac{\xi \pm \sqrt{\xi ^2-m^2}}M.
\end{equation}
Using the relations (\ref{k11}), (\ref{eq10}) one can easily check 
\begin{equation}  \label{eq14}
1\geq x_{+}\geq \frac mM\geq x_{-}\geq \frac{m^2}{M^2},\qquad E_{\max }=%
\frac{M^2+m^2}{2M}\geq \xi \geq m.
\end{equation}
First let us insert $x_{+}$ into (\ref{eq12}) 
\begin{equation}  \label{eq16}
F\left( \frac{\xi +\sqrt{\xi ^2-m^2}}M\right) =2\pi M\int_\xi ^{E_{\max
}}G(p_0)p_0dp_0.
\end{equation}
Differentiation in respect to $\xi $ gives 
\begin{equation}  \label{eq17}
G(\xi )=-\frac 1{2\pi M^2}F^{\prime }\left( \frac{\xi +\sqrt{\xi ^2-m^2}}%
M\right) \left( \frac 1\xi +\frac 1{\sqrt{\xi ^2-m^2}}\right) .
\end{equation}
Now we integrate the density $G(p_0)$ over angular variables obtaining 
\begin{equation}  \label{eq18}
P(p_0)dp_0\equiv \int_\Omega G(p_0)d^3p=4\pi G(p_0)p_0\sqrt{p_0^2-m^2}dp_0
\end{equation}
and after inserting into (\ref{eq17}) we get 
\begin{equation}  \label{eq19}
P(p_0)dp_0=-2F^{\prime }\left( \frac{p_0+\sqrt{p_0^2-m^2}}M\right) \frac{p_0+%
\sqrt{p_0^2-m^2}}M\frac{dp_0}M.
\end{equation}
Second root $x_{-}$ gives very similar result 
\begin{equation}  \label{eq19a}
P(p_0)dp_0=+2F^{\prime }\left( \frac{p_0-\sqrt{p_0^2-m^2}}M\right) \frac{p_0-%
\sqrt{p_0^2-m^2}}M\frac{dp_0}M.
\end{equation}
From the definition 
\begin{equation}  \label{eq19b}
x_{\pm }\equiv \frac{p_0\pm \sqrt{p_0^2-m^2}}M,
\end{equation}
the useful relations easily follow 
\begin{equation}  \label{eq19c}
x_{+}x_{-}=\frac{m^2}{M^2},\qquad x_{+}+x_{-}=\frac{2p_0}M,\qquad
x_{+}-x_{-}=\frac{2\sqrt{p_0^2-m^2}}M,
\end{equation}
\begin{equation}  \label{eq19d}
\frac{dp_0}M=\frac 12(1-\frac{m^2}{M^2x_{\pm }^2})dx_{\pm },\qquad \frac{%
dx_{+}}{x_{+}}=-\frac{dx_{-}}{x_{-}}.
\end{equation}
Now, the equations (\ref{eq19}), (\ref{eq19a}) can be joined 
\begin{equation}  \label{eq20}
P(p_0)=\mp \frac 2MF^{\prime }(x_{\pm })x_{\pm }.
\end{equation}
How to understand the two different partial intervals (\ref{eq14}) of $x$
give independently the complete distribution $P(p_0)$ in Eq. (\ref{eq20})?
It is due to the fact that e.g. $x_{-}$ represents in the integral (\ref{eq4}%
) the region 
\begin{equation}  \label{eq22}
\frac{\sqrt{p_1^2+p_T^2+m^2}+p_1}M=x_{-}\leq \frac mM.
\end{equation}
given by the paraboloid 
\begin{equation}  \label{eq23}
p_T^2\leq 2m\left| p_1\right| ,\qquad p_1\leq 0,
\end{equation}
containing complete information about $G(p_0)$ which is spherically
symmetric. The similar argument is valid for $x_{+}$ representing the rest
of sphere. The Eqs. (\ref{eq19}), (\ref{eq19a}) imply the similarity of $%
F(x) $ in both intervals 
\begin{equation}  \label{eq24}
\frac{F^{\prime }(x_{+})x_{+}}{F^{\prime }(x_{-})x_{-}}=-1,
\end{equation}
which with the use of second relation (\ref{eq19d}) can be easily shown to
be equivalent to 
\begin{equation}  \label{eq24a}
F(x_{+})=F(x_{-}).
\end{equation}
The relation (\ref{eq20}) implies the distribution function $F(x)$ should be
increasing for $(m/M)^2<x<m/M$ and decreasing for $m/M<x<1$ e.g. as shown in
Fig. \ref{yy2}. Now let us calculate the following integrals. 

\textit{The total number $N$ of partons:} 
\begin{equation}  \label{eq25}
N=\int_m^{E_{\max }}P(p_0)dp_0=-\int_{m/M}^1F^{\prime }(x_{+})(x_{+}-\frac{%
m^2}{M^2x_{+}})dx_{+}
\end{equation}
\begin{equation*}
=-\int_{m/M}^1F^{\prime }(x_{+})x_{+}dx_{+}+\int_{m/M}^1F^{\prime
}(x_{+})x_{-}dx_{+}.
\end{equation*}
The last integral can be modified with the use of (\ref{eq19d}), (\ref{eq24}%
) 
\begin{equation}  \label{eq25a}
\int_{m/M}^1F^{\prime }(x_{+})x_{-}dx_{+}=-\int_{m/M}^1F^{\prime
}(x_{-})x_{-}^2\frac{dx_{+}}{x_{+}}=\int_{m/M}^{m^2/M^2}F^{\prime
}(x_{-})x_{-}dx_{-}.
\end{equation}
Then integration by parts gives 
\begin{equation}  \label{eq25b}
N=-\int_{m^2/M^2}^1F^{\prime }(x)xdx=\int_{m^2/M^2}^1F(x)dx.
\end{equation}
\textit{The total energy $E$ of partons:} 
\begin{equation}  \label{eq26}
E=\int_m^{E_{\max }}P(p_0)p_0dp_0=-\int_{m/M}^1F^{\prime }(x_{+})(x_{+}-%
\frac{m^2}{M^2x_{+}})\frac M2(x_{+}+x_{-})dx_{+}=
\end{equation}
\begin{equation*}
=-\frac M2\int_{m/M}^1F^{\prime }(x_{+})(x_{+}^2-x_{-}^2)dx_{+}.
\end{equation*}
A similar procedure as for $N$ then gives the result 
\begin{equation}  \label{eq27}
E=-\frac M2\int_{m^2/M^2}^1F^{\prime }(x)x^2dx=M\int_{m^2/M^2}^1F(x)xdx.
\end{equation}
Therefore, the both descriptions based either on the IMF variable $x$ or the
parton energy $p_0$ in the LAB give the consistent results on the total
number of partons and the fraction of energy carried by the partons. Let us
remark, a model based on the spherically symmetric Gaussian distribution of
the parton momenta in the hadron rest frame has been recently proposed in %
\cite{ing}.

\subsection{Structure functions $F_2, F_1$}

An important connection between the structure and distribution functions can
be within QPM derived by a few (equivalent) ways, see e.g. textbooks \cite%
{fey}-\cite{ait}. In this section we shall consider the electromagnetic
unpolarized structure functions assuming quasifree partons with spin 1/2.
The general form of cross section for the scattering \textit{electron +
proton} and \textit{electron + point like, Dirac particle} can be written 
\begin{equation}
d\sigma (e^{-}+p)=\frac{e^{4}}{q^{4}}\frac{1}{4\sqrt{(kP)^{2}-m_{e}^{2}M^{2}}%
}K^{\alpha \beta }W_{\alpha \beta }4\pi M\frac{d^{3}k^{\prime }}{%
2k_{0}^{\prime }(2\pi )^{3}},  \label{sf00}
\end{equation}%
\begin{equation}
d\sigma (e^{-}+l)=\frac{e^{4}}{q^{4}}\frac{1}{4\sqrt{%
(kp)^{2}-m_{e}^{2}m_{l}^{2}}}K^{\alpha \beta }L_{\alpha \beta }2\pi \delta
((p+q)^{2}-m^{2})\frac{d^{3}k^{\prime }}{2k_{0}^{\prime }(2\pi )^{3}},
\label{sf01}
\end{equation}%
where the electron tensor has the standard form 
\begin{equation}
K^{\alpha \beta }=2(k^{\alpha }k^{\prime \beta }+k^{\prime \alpha }k^{\beta
}+g^{\alpha \beta }\frac{q^{2}}{2})  \label{sf02}
\end{equation}%
and the remaining hadron and lepton tensors $W_{\alpha \beta },$ $L_{\alpha
\beta }$ can be written in the ''reduced'' shape 
\begin{equation}
W_{\alpha \beta }=\frac{P_{\alpha }P_{\beta }}{M^{2}}W_{2}-g_{\alpha \beta
}W_{1},  \label{sf03}
\end{equation}%
\begin{equation}
L_{\alpha \beta }=4p_{\alpha }p_{\beta }-2g_{\alpha \beta }pq.  \label{sf04}
\end{equation}%
General assumption that the scattering on proton is realized via scattering
on the partons implies 
\begin{equation}
d\sigma (e^{-}+p)=\int F(\xi )d\sigma (e^{-}+l)d\xi ,  \label{sf05}
\end{equation}%
where $F(\xi )$ is a probabilistic function describing distribution of
partons according to some parameter(s) $\xi .$ Now, if $F(\xi )$ is
substituted by the usual distribution function and we assume 
\begin{equation}
p_{\alpha }=\xi P_{\alpha }.  \label{sf2}
\end{equation}%
then it is obvious, that Eq. (\ref{sf05}) after inserting from Eqs. (\ref%
{sf00}) and (\ref{sf01}) will be satisfied provided that 
\begin{equation}
P_{\alpha }P_{\beta }\frac{W_{2}}{M^{2}}-g_{\alpha \beta }W_{1}=\frac{1}{M}%
\int \frac{F(\xi )}{\xi }(2\xi ^{2}P_{\alpha }P_{\beta }-g_{\alpha \beta
}\,\xi Pq)\delta ((\xi P+q)^{2}-m^{2})d\xi .  \label{sf3}
\end{equation}%
For simplicity in this equation, and anywhere in this section, the weighting
by the parton charges is omitted. In fact the Eq. (\ref{sf3}) is just a
master equation in \cite{fey}(lesson 27, Eq. (27.4)), from which the known
relations follow: 
\begin{equation}
2MW_{1}(q^{2},\nu )=\frac{F_{2}(x)}{x},\qquad xF(x)=F_{2}(x)\equiv \nu
W_{2}(q^{2},\nu ),\qquad x\equiv \frac{-q^{2}}{2M\nu }.  \label{sf4}
\end{equation}%
Here, let us point out, this result is based on the approximation (\ref{sf2}%
), which is currently accepted in IMF. In fact, relation (\ref{sf2}) in the
covariant formulation is equivalent to the assumption, that the partons are
static with respect to the nucleon, therefore there are suppressed not only
the transversal momenta, but also longitudinal ones. In the LAB this
relation implies $\xi =m/M$, so in the case of our quasifree partons,
corresponding distribution function reflects rather distribution of the
parton masses.

Before repeating the above procedure for our distribution $G(p_0)d^3p$ in
LAB, one has correctly account for the flux factor corresponding to partons
moving inside the proton volume with velocity $\vec v=\vec p/p_0$. If this
velocity has the opposite direction to the probing electron, then after
passing through the whole subset $G(p_0)d^3p$ the electron has not still
reached backward boundary of the proton, where meanwhile the new partons
appeared. And on contrary, if the velocity of subset has the same direction
as the electron, then not all of these partons have the same chance to meet
this electron. Namely, the partons close to the backward boundary are
excluded from the game sooner than the electron reaches them.
Quantitatively, in a subset of partons $G(p_0)d^3p,$ the number of partons
limited by the proton volume and having chance to meet the probing electron
will be 
\begin{equation}  \label{sf07}
dN=(1-v_1/v_e)G(p_0)d^3p,
\end{equation}
where $v_1=p_1/p_0$ is the component of parton velocity in the direction of
the passing electron, $v_e\approx -1$ is the electron velocity, if we assume
electron momentum $k=(k_0,-k_0,0,0)$. Therefore, one can put 
\begin{equation}  \label{sfa07}
d\sigma (e+p)=\int (1+p_1/p_0)G(p_0)d\sigma (e+q)d^3p.
\end{equation}
We neglect the electron mass, so inserting from Eqs. (\ref{sf00}),(\ref{sf01}%
) gives 
\begin{equation}  \label{ff1}
K^{\alpha \beta }W_{\alpha \beta }=K^{\alpha \beta }\frac{kP}{2M}\int \frac{%
(1+p_1/p_0)}{kp}G(p_0)L_{\alpha \beta }\delta ((p+q)^2-m^2)d^3p.
\end{equation}
The flux factors expressed in the proton rest frame $kP=k_0M,kp=k_0(p_0+p_1)$
and tensors $W_{\alpha \beta },L_{\alpha \beta }$ from Eqs. (\ref{sf03}),(%
\ref{sf04}) inserted to the last equation give 
\begin{equation*}
K^{\alpha \beta }\left\{ P_\alpha P_\beta \frac{W_2}{M^2}-g_{\alpha \beta
}W_1\right\}
\end{equation*}
\begin{equation}  \label{t1}
=K^{\alpha \beta }\left\{ \int G(p_0)(2p_\alpha p_\beta -g_{\alpha \beta
}pq)\delta ((p+q)^2-m^2)\frac{d^3p}{p_0}\right\} .
\end{equation}
The last equation can be rewritten 
\begin{equation}  \label{t2}
K^{\alpha \beta }(l_{\alpha \beta }-r_{\alpha \beta })=0,
\end{equation}
where $l,r$ are corresponding tensors $\left\{ ...\right\} $ in the l.h.s.
and r.h.s. of Eq. (\ref{t1}). Since the tensor $K^{\alpha \beta }$ obeys the
current conservation 
\begin{equation}  \label{t3}
q_\alpha K^{\alpha \beta }=q_\beta K^{\alpha \beta }=0,
\end{equation}
Eq. (\ref{t2}) will be satisfied, if the difference $(l-r)_{\alpha \beta }$
has the form $A(P_\alpha q_\beta +P_\beta q_\alpha )+Bq_\alpha q_\beta .$ In
this way we get the tensor equation 
\begin{equation}  \label{sf5}
P_\alpha P_\beta \frac{W_2}{M^2}-g_{\alpha \beta }W_1+A(P_\alpha q_\beta
+P_\beta q_\alpha )+Bq_\alpha q_\beta
\end{equation}
\begin{equation*}
=\int \frac{G(p_0)}{p_0}(2p_\alpha p_\beta -g_{\alpha \beta }\,pq)\delta
((p+q)^2-m^2)d^3p,\qquad p_0=\sqrt{m^2+p_1^2+p_2^2+p_3^2,}
\end{equation*}
for which (\ref{sf2})\textit{\ is not required}. One can prove, that the
tensors $W,L$ in the form satisfying (\ref{t3}), would obey Eq. (\ref{sf5})
in which $A=B=0.$ The terms with the functions $A$ and $B$ do not contribute
to the cross section. Also let us note, the velocity correction similar to (%
\ref{sf07}) was not used in the Eq. (\ref{sf3}) since due to (\ref{sf2}) the
all partons in the applied approach have the same velocity as the proton
i.e. $v_1=0$.

Now the contracting of (\ref{sf5}) with tensors $g^{\alpha \beta },q^\alpha
q^\beta ,P^\alpha P^\beta ,P^\alpha q^\beta $ gives as a result the set of
four equations 
\begin{equation}  \label{r1}
W_2-4W_1+2M\nu (A-xB)=\frac 1{M\nu }\int \frac{G(p_0)}{p_0}[m^2-2Mx\nu
]\delta \left( \frac{pq}{M\nu }-x\right) d^3p,
\end{equation}
\begin{equation}  \label{r2}
\frac \nu {2Mx}W_2+W_1-2M\nu (A-xB)=\frac 1{M\nu }\int \frac{G(p_0) }{p_0}%
[Mx\nu ]\delta \left( \frac{pq}{M\nu }-x\right) d^3p,
\end{equation}
\begin{equation}  \label{r3}
W_2-W_1+\nu (2MA+\nu B)=\frac 1{M\nu }\int \frac{G(p_0)}{p_0}[p_0^2-\frac{%
Mx\nu }2]\delta \left( \frac{pq}{M\nu }-x\right) d^3p,
\end{equation}
\begin{equation}  \label{r4}
W_2-W_1+(M\nu -2M^2x)A-2Mx\nu B
\end{equation}
\begin{equation*}
=\frac 1{M\nu }\int \frac{G(p_0)}{p_0}[p_0Mx-\frac{Mx\nu }2]\delta \left( 
\frac{pq}{M\nu }-x\right) d^3p,
\end{equation*}
in which the $\delta -$function from the integral (\ref{sf5}) is expressed 
\begin{equation}  \label{sf6}
\delta ((p+q)^2-m^2)=\delta (2pq+q^2)
\end{equation}
\begin{equation*}
=\delta \left( 2M\nu \left( \frac{pq}{M\nu }-\frac{Q^2}{2M\nu }\right)
\right) =\frac 1{2M\nu }\delta \left( \frac{pq}{M\nu }-x\right) .
\end{equation*}
If we define 
\begin{equation}  \label{sf11}
V_j(x)\equiv \int G(p_0)\left( \frac{p_0}M\right) ^j\delta \left( \frac{pq}{%
M\nu }-x\right) d^3p,
\end{equation}
then the solution of the set (\ref{r1})-(\ref{r4}) reads 
\begin{equation}  \label{r5}
2MW_1=\frac \nu {2Mx+\nu }\cdot \left\{ V_{-1}(x)\left[ x-\frac M\nu \left( 
\frac{m^2}{M^2}-x^2\right) -2\frac{m^2x}{\nu ^2}\right] \right.
\end{equation}
\begin{equation*}
\left. +V_0(x)\frac{2Mx}\nu +V_1(x)\frac{2M^2x}{\nu ^2}\right\} ,
\end{equation*}
\begin{equation}  \label{r6}
\nu W_2=x\left( \frac \nu {2Mx+\nu }\right) ^2\cdot \left\{ V_{-1}(x)\left[
x-\frac M\nu \left( \frac{m^2}{M^2}+x^2\right) -2\frac{m^2x}{\nu ^2}\right]
\right.
\end{equation}
\begin{equation*}
\left. +V_0(x)\frac{6Mx}\nu +V_1(x)\frac{6M^2x}{\nu ^2}\right\} ,
\end{equation*}
\begin{equation}  \label{r7}
\nu ^2MA=-\left( \frac \nu {2Mx+\nu }\right) ^2\cdot \left\{ V_{-1}(x)\left[
\frac 12\left( \frac{m^2}{M^2}+3x^2\right) +\frac{m^2x}{M\nu }\right] \right.
\end{equation}
\begin{equation*}
\left. -V_0(x)2x\left( 1-\frac{Mx}\nu \right) -V_1(x)\frac{3Mx}\nu \right\} ,
\end{equation*}
\begin{equation}  \label{r8}
\nu ^3B=\left( \frac \nu {2Mx+\nu }\right) ^2\cdot \left\{ V_{-1}(x)\left[
\frac 12\left( \frac{m^2}{M^2}+3x^2\right) +\frac{m^2x}{M\nu }\right] \right.
\end{equation}
\begin{equation*}
\left. -V_0(x)3x+V_1(x)\left( 1-\frac{Mx}\nu \right) \right\} .
\end{equation*}
For next discussion we assume $\nu \gg M$, then 
\begin{equation}  \label{r9}
\nu W_2\equiv F_2(x)=x^2V_{-1}(x),\qquad MW_1\equiv F_1(x)=\frac x2V_{-1}(x),
\end{equation}
so it is obvious the Callan-Gross relation $2xF_1=F_2$ holds in this
approximation.

In the next step, with the use of the approximation (\ref{kcj10}), we
express the integrals (\ref{sf11}): 
\begin{equation}  \label{r10}
V_j(x)=\int G(p_0)\left( \frac{p_0}M\right) ^j\delta \left( \frac{p_0+p_1}%
M-x\right) d^3p.
\end{equation}
This relation with the use of (\ref{eq4}),(\ref{eq20}) implies 
\begin{equation}  \label{r11}
\left( \frac{p_0}M\right) ^jP(p_0)=\mp \frac 2MV_j^{\prime }(x_{\pm })x_{\pm
},
\end{equation}
where $x_{\pm }$ is defined in (\ref{eq19b}). The relations (\ref{r11}) and (%
\ref{eq19c}) give 
\begin{equation}  \label{r12}
\frac{V_j^{\prime }(x)}{V_k^{\prime }(x)}=\left( \frac{p_0}M\right)
^{j-k}=\left( \frac{x_{+}+x_{-}}2\right) ^{j-k}=\left( \frac x2+ \frac{x_0^2%
}{2x}\right) ^{j-k},\qquad x_0=\frac mM.
\end{equation}
In the previous section we have shown such functions as Eq. (\ref{r10}) obey
the relation (\ref{eq24a}), which means in particular, that the functions
have a maximum at $x_0$ and vanish for $x\leq x_0^2$. Therefore the same
statement is valid also for functions $F_2/x^2$ and $F_1/x$ from Eq. (\ref%
{r9}) 
\begin{equation}  \label{r121}
\frac{F_2(x_{+})}{x_{+}^2}=\frac{F_2(x_{-})}{x_{-}^2},\qquad \frac{F_1(x_{+})%
}{x_{+}}=\frac{F_1(x_{-})}{x_{-}}.
\end{equation}
This means that the structure functions of our idealized nucleon have the
maximum at $x_0$ or higher.

Further, our considerations have started to move in previous section from
the distribution function $F(x)$ for which we have obtained relation (\ref%
{eq20}). The combination of this equation with (\ref{r9}), (\ref{r11}) and (%
\ref{eq19c}) gives 
\begin{equation}  \label{r122}
P(p_0)=-\frac 1M\left( \frac{F_2(x)}{x^2}\right) ^{\prime
}(x^2+x_0^2),\qquad x=\frac{p_0+\sqrt{p_0^2-m^2}}M,
\end{equation}
\begin{equation}  \label{r13}
F^{\prime }(x)=\frac 12\left( \frac{F_2(x)}{x^2}\right) ^{\prime }\left( x+%
\frac{x_0^2}x\right) .
\end{equation}
How do we compare the last equation with the standard relation (\ref{sf4})
for $F$ and $F_2$? As we have already told, the standard approach (\ref{sf3}%
) would be exact in the case when the partons are static with respect to the
nucleon, i.e. when $x=m/M$. The Eq. (\ref{sf5}) itself is more exact, but in
further procedure we assume the masses of all the partons in the considered
subset being equal. Therefore for a comparison let us consider first the
extreme scenario when the parton distribution functions $F(x)$ and $P(p_0)$
are [(see Eq. (\ref{eq20})] rather narrowly peaked around the points $%
x_0=m/M $ and $p_0=m$. Then for $x\approx x_0$ Eq. (\ref{r13}) gives 
\begin{equation}  \label{r14}
F^{\prime }(x)=\frac 12\left( \frac{F_2(x)}{x^2}\right) ^{\prime }\left( x+%
\frac{x_0^2}x\right) \simeq \frac 12\frac{F_2^{\prime }(x)}{x_0^2}(x_0+x_0)=%
\frac{F_2^{\prime }(x)}{x_0}
\end{equation}
from which the second relation (\ref{sf4}) follows as a limiting case of (%
\ref{r13}) 
\begin{equation}  \label{sf33}
x_0F(x_0)\approx F_2(x_0).
\end{equation}
Now, in the case when the distribution functions are broad, the exact
validity of Eq. (\ref{sf3}) again requires static partons, therefore the
corresponding distribution function represents also a spectrum of masses.
But then obviously the above procedure for a single $m$ can be repeated with
spectrum of masses $F(x_0)$ giving in the result instead of Eq. (\ref{sf33})
the relation 
\begin{equation}  \label{r15}
\int x_0F(x_0)\delta (x-x_0)dx_0=\int F_2(x_0)\delta (x-x_0)dx_0,
\end{equation}
which implies 
\begin{equation}  \label{r16}
xF(x)=F_2(x).
\end{equation}
In this sense the standard approach based on Eq. (\ref{sf3}) can be
understood as a limiting case (static partons) of that based on Eq. (\ref%
{sf5}).

\subsection{Spin structure functions $g_1, g_2$}

\label{spin}

In the previous section the master equation (\ref{sf5}) has been based on
the standard symmetric tensors (\ref{sf03}) and (\ref{sf04}) corresponding
to the unpolarized DIS. After introduction the spin terms into both the
tensors (see e.g. \cite{fey}, Eqs. (33.9), (33.10)) the master equation
reads 
\begin{equation}
P_{\alpha }P_{\beta }\frac{W_{2}}{M^{2}}-g_{\alpha \beta }W_{1}+i\epsilon
_{\alpha \beta \lambda \sigma }q^{\lambda }\left[ S^{\sigma
}MG_{1}+(PqS^{\sigma }-SqP^{\sigma })\frac{G_{2}}{M}\right]  \label{ss1}
\end{equation}%
\begin{equation*}
+A(P_{\alpha }q_{\beta }+P_{\beta }q_{\alpha })+Bq_{\alpha }q_{\beta }=\int
G(p_{0})(2p_{\alpha }p_{\beta }-g_{\alpha \beta }\,pq)\delta
((p+q)^{2}-m^{2})\frac{d^{3}p}{p_{0}}
\end{equation*}%
\begin{equation*}
+i\epsilon _{\alpha \beta \lambda \sigma }q^{\lambda }\int
H(p_{0})mw^{\sigma }\delta ((p+q)^{2}-m^{2})\frac{d^{3}p}{p_{0}},
\end{equation*}%
where $G$ and $H$ are related to the polarized quark distributions 
\begin{equation}
G(p_{0})=\sum_{j}e_{j}^{2}(h_{j}^{\uparrow }(p_{0})+h_{j}^{\downarrow
}(p_{0})),  \label{ss2}
\end{equation}%
\begin{equation}
H(p_{0})=\sum_{j}e_{j}^{2}(h_{j}^{\uparrow }(p_{0})-h_{j}^{\downarrow
}(p_{0}))  \label{ss3}
\end{equation}%
and the spin polarization vectors satisfy 
\begin{equation}
S_{\mu }S^{\mu }=w_{\mu }w^{\mu }=-1,\qquad S_{\mu }P^{\mu }=w_{\mu }p^{\mu
}=0.  \label{ss4}
\end{equation}%
The Eq. (\ref{ss1}) requires for the spin terms 
\begin{equation}
S^{\sigma }MG_{1}+(PqS^{\sigma }-SqP^{\sigma })\frac{G_{2}}{M}=\frac{m}{%
2M\nu }\int \frac{H(p_{0})}{p_{0}}w^{\sigma }\delta \left( \frac{pq}{M\nu }%
-x\right) d^{3}p,  \label{ss5}
\end{equation}%
where we use for the $\delta -$function the relation (\ref{sf6}).

Now, let us consider first a extremely simple scenario (in LAB) assuming the
following.

\noindent 1) To the function $H$ in Eq. (\ref{ss3}) only the valence quark
term contributes.

\noindent 2) Momenta distributions have the same (spherically symmetric)
shape for $u$ and $d$ quarks 
\begin{equation}  \label{ss6}
h_d(p_0)=\frac 12h_u(p_0)\equiv h(p_0)
\end{equation}
and both the quarks have the same mass $m$.

\noindent 3) All the three quarks contribute to the proton spin equally 
\begin{equation}  \label{ss7}
h_d^{\uparrow }-h_d^{\downarrow }=\frac 12(h_u^{\uparrow }-h_u^{\downarrow
})\equiv \Delta h(p_0)=\frac 13h(p_0),\qquad p_0=\sqrt{m^2+p_1^2+p_2^2+p_3^2}%
,
\end{equation}
i.e. in a first step we ignore constraints due to axial vector current
operators on the spin contribution from different flavors. Since all the
three quarks are assumed to give the proton spin, the last equation implies 
\begin{equation}  \label{ss8}
3\int \Delta h(p_0)d^3p=1.
\end{equation}
The combination with (\ref{ss3}) gives 
\begin{equation}  \label{ss9}
H(p_0)=2\frac 49\Delta h(p_0)+\frac 19\Delta h(p_0)=\Delta h(p_0)
\end{equation}
and 
\begin{equation}  \label{ss10}
\int H(p_0)d^3p=\frac 13.
\end{equation}

Now, let us assume the proton is polarized in the direction of the collision
axis (coordinate one), then Eqs. (\ref{ss4}),(\ref{ss5}) require for the
proton at rest 
\begin{equation}
S=(0,1,0,0)  \label{ss11}
\end{equation}%
and for the quark with four-momentum $p$, 
\begin{equation}
w=\left( \frac{p_{1}}{\sqrt{p_{0}^{2}-p_{1}^{2}}},\frac{p_{0}}{\sqrt{%
p_{0}^{2}-p_{1}^{2}}},0,0\right) .  \label{ss12}
\end{equation}%
More rigorous derivation of this form of the quark polarization vector,
which is based on the requirement of the relativistic covariance, is done in
the next section. The contracting of Eq. (\ref{ss5}) with $P_{\sigma }$ and $%
S_{\sigma }$ (or equivalently, simply taking $\sigma =0,1)$ gives the
equations 
\begin{equation}
q_{1}G_{2}=\frac{m}{2M\nu }\int \frac{H(p_{0})}{p_{0}}\frac{p_{1}}{\sqrt{%
p_{0}^{2}-p_{1}^{2}}}\delta \left( \frac{pq}{M\nu }-x\right) d^{3}p,
\label{ss13}
\end{equation}%
\begin{equation}
MG_{1}+\nu G_{2}=\frac{m}{2M\nu }\int \frac{H(p_{0})}{p_{0}}\frac{p_{0}}{%
\sqrt{p_{0}^{2}-p_{1}^{2}}}\delta \left( \frac{pq}{M\nu }-x\right) d^{3}p.
\label{ss14}
\end{equation}%
In the next step we apply the approximations from the Eqs. (\ref{ka10}) and (%
\ref{kcj10}) 
\begin{equation}
q_{1}\simeq -\nu ,\qquad \frac{pq}{M\nu }\simeq \frac{p_{0}+p_{1}}{M}.
\label{ss15}
\end{equation}%
Let us note, the negative sign in the first relation is connected with the
choice of the lepton beam direction giving the Eq. (\ref{kcj10}). The
opposite choice should give 
\begin{equation}
q_{1}\simeq +\nu ,\qquad \frac{pq}{M\nu }\simeq \frac{p_{0}-p_{1}}{M}
\label{ss16}
\end{equation}%
and one can check the both alternatives result in the equal pairs $%
G_{1},G_{2}$, which read 
\begin{equation}
2g_{1}(x)\equiv 2M^{2}\nu G_{1}=m\int \frac{H(p_{0})}{p_{0}}\frac{p_{0}+p_{1}%
}{\sqrt{p_{0}^{2}-p_{1}^{2}}}\delta \left( \frac{p_{0}+p_{1}}{M}-x\right)
d^{3}p,  \label{ss17}
\end{equation}%
\begin{equation}
2g_{2}(x)\equiv 2M\nu ^{2}G_{2}=-m\int \frac{H(p_{0})}{p_{0}}\frac{p_{1}}{%
\sqrt{p_{0}^{2}-p_{1}^{2}}}\delta \left( \frac{p_{0}+p_{1}}{M}-x\right)
d^{3}p.  \label{ss18}
\end{equation}%
Let us remark, the integration of Eqs. (\ref{ss13}) and (\ref{ss18}) over $x$
gives on r.h.s. the integral 
\begin{equation}
\int \frac{H(p_{0})}{p_{0}}\frac{p_{1}}{\sqrt{p_{0}^{2}-p_{1}^{2}}}d^{3}p=0,
\label{ssa18}
\end{equation}%
which is zero due to spherical symmetry. Therefore in this approach the
first moment of $g_{2}(x)$ is zero as well. Now we shall pay attention
particularly to the function $g_{1}$, which can be rewritten 
\begin{equation}
2g_{1}(x)=\frac{x_{0}}{3}\int h(p_{0})\frac{M}{p_{0}}\sqrt{\frac{p_{0}+p_{1}%
}{p_{0}-p_{1}}}\delta \left( \frac{p_{0}+p_{1}}{M}-x\right) d^{3}p,\qquad
x_{0}=\frac{m}{M}.  \label{ss19}
\end{equation}%
What our assumptions 1)-3) do mean in the language of the standard IMF
approach? In the previous section we have suggested that our approach is
equivalent to the standard one [based on the approximation (\ref{sf2})], for
the static quarks described by the distribution function $h(p_{0})$ sharply
peaked around $m$. In such a case the last equation for $p_{0}\approx m,$ $%
p_{1}\approx 0$ after combining with (\ref{ss3}) and (\ref{eq4}) gives 
\begin{equation}
2g_{1}(x)=\int \sum_{j}e_{j}^{2}\left( h_{j}^{\uparrow
}(p_{0})-h_{j}^{\downarrow }(p_{0})\right) \delta \left( \frac{p_{0}+p_{1}}{M%
}-x\right) d^{3}p  \label{ss20}
\end{equation}%
\begin{equation*}
=\sum_{j}e_{j}^{2}\left( f_{j}^{\uparrow }(x)-f_{j}^{\downarrow }(x)\right) ,
\end{equation*}%
where $f_{j}(x)$ are corresponding distribution functions in the IMF, so in
this limiting case our spin equation (\ref{ss19}) is again identical to the
standard one, see Eq. (33.14) in \cite{fey}. The last equation can be in our
simplified scenario rewritten 
\begin{equation}
2g_{1}(x)=\frac{1}{3}\int h(p_{0})\delta \left( \frac{p_{0}+p_{1}}{M}%
-x\right) d^{3}p=\frac{1}{3}f(x)=\frac{F_{2val}(x)}{3x}.  \label{ss21}
\end{equation}%
This relation says, what our assumptions 1)-3) mean in the terms of the IMF
approach, in particular we obtain 
\begin{equation}
\Gamma _{IMF}\equiv \int g_{1}(x)dx=\frac{1}{6}\int f(x)dx=\frac{1}{6}.
\label{ss22}
\end{equation}%
Further, in accordance with (\ref{r10}) let us denote 
\begin{equation}
V_{j}(x)\equiv \int h(p_{0})\left( \frac{p_{0}}{M}\right) ^{j}\delta \left( 
\frac{p_{0}+p_{1}}{M}-x\right) d^{3}p,  \label{ss23}
\end{equation}%
then (\ref{r9}) and (\ref{ss21}) give 
\begin{equation}
2g_{1}(x)=\frac{xV_{-1}(x)}{3},\qquad \Gamma _{IMF}=\frac{1}{6}\int
xV_{-1}(x)dx.  \label{ss24}
\end{equation}

So, our Eq. (\ref{ss19}) in the limit case of the static quarks coincides
with the standard IMF approach, but what this equation implies for the
nonstatic quarks? Let us calculate the first moment of our $g_1$: 
\begin{equation}  \label{ss25}
\Gamma _{lab}=\frac{x_0}6\int \int h(p_0)\frac M{p_0}\sqrt{\frac{p_0+p_1}{%
p_0-p_1}}\delta \left( \frac{p_0+p_1}M-x\right) d^3pdx.
\end{equation}
Due to the $\delta -$ function, the square root term in the integral can be
rewritten 
\begin{equation}  \label{ss26}
\sqrt{\frac{p_0+p_1}{p_0-p_1}}=\sqrt{\frac{Mx}{2p_0-Mx}}
\end{equation}
\begin{equation*}
=\sqrt{\frac{Mx}{2p_0}}\left( 1-\frac{Mx}{2p_0}\right) ^{-1/2}=\left( \frac{%
Mx}{2p_0}\right) ^{1/2}\sum_{j=0}^\infty {\binom{{-\frac 12}}j}(-1)^j\left( 
\frac{Mx}{2p_0}\right) ^j
\end{equation*}
and using Eq. (\ref{ss23}) the integral is, correspondingly 
\begin{equation}  \label{ss27}
\Gamma _{lab}=\frac{x_0}6\int \sum_{j=0}^\infty {\binom{{-\frac 12}}j}%
(-1)^jV_{-j-3/2}(x)\left( \frac x2\right) ^{j+1/2}dx.
\end{equation}
The integration by parts combined with the relations (\ref{r12}) gives%
\begin{equation*}
\int V_{-j-3/2}(x)\left( \frac x2\right) ^{j+1/2}dx
\end{equation*}
\begin{equation*}
=\int V_{-j-3/2}^{\prime }(x)\frac{2\left( x/2\right) ^{j+3/2}}{j+3/2}%
dx=\int V_0^{\prime }(x)\left( \frac x2+\frac{x_0^2}{2x}\right) ^{-j-3/2} 
\frac{2\left( x/2\right) ^{j+3/2}}{j+3/2}dx
\end{equation*}
\begin{equation*}
=\int V_0^{\prime }(x)\frac 2{j+3/2}\left( \frac 1{1+x_0^2/x^2}\right)
^{j+3/2}dx
\end{equation*}
\begin{equation*}
=\int V_0(x)2\left( \frac 1{1+x_0^2/x^2}\right) ^{j+1/2}\frac{2x_0^2/x^3}{%
\left( 1+x_0^2/x^2\right) ^2}dx.
\end{equation*}
If we denote $t\equiv x_0^2/x^2$ and $z\equiv 1/(1+t^2)$ then Eq. (\ref{ss27}%
) can be rewritten as%
\begin{equation*}
\Gamma _{lab}=\frac 16\int V_0(x)4t^3z^{5/2}\sum_{j=0}^\infty {\binom{{%
-\frac 12}}j}(-1)^jz^jdx=\frac 16\int V_0(x)4t^3z^2\sqrt{\frac z{1-z}}dx,
\end{equation*}
which implies 
\begin{equation}  \label{ss28}
\Gamma _{lab}=\frac 16\int_{x_0^2}^1\frac{4x_0^2/x^2}{(1+x_0^2/x^2)^2}%
V_0(x)dx.
\end{equation}
Simultaneously, since 
\begin{equation*}
\int_{x_0^2}^1V_0(x)dx=-\int_{x_0^2}^1xV_0^{\prime
}(x)dx=-\int_{x_0^2}^1xV_{-1}^{\prime }(x)\left( \frac x2+\frac{x_0^2}{2x}%
\right) dx
\end{equation*}
\begin{equation*}
=-\int_{x_0^2}^1V_{-1}^{\prime }(x)\left( \frac{x^2}2+\frac{x_0^2}2\right)
dx=\int_{x_0^2}^1V_{-1}(x)xdx,
\end{equation*}
the integral (\ref{ss24}) can be rewritten as 
\begin{equation}  \label{ss29}
\Gamma _{IMF}=\frac 16\int_{x_0^2}^1V_0(x)dx.
\end{equation}
Let us express the last integral as%
\begin{equation*}
\int_{x_0^2}^1V_0(x)dx=\int_{x_0^2}^{x_0}V_0(x)dx+\int_{x_0}^1V_0(x)dx
\end{equation*}
and modify the first integral on r.h.s. using substitution $y=x_0^2/x$%
\begin{equation*}
\int_{x_0^2}^{x_0}V_0(x)dx=\int_{x_0}^1V_0\left( \frac{x_0^2}y\right) \frac{%
x_0^2}{y^2}dy.
\end{equation*}
Now let us recall the general shape of the functions (\ref{ss23}) obeying
Eq. (\ref{eq24a}), which implies%
\begin{equation*}
V_0\left( \frac{x_0^2}y\right) =V_0(y),
\end{equation*}
so instead of Eq. (\ref{ss29}) one can write 
\begin{equation}  \label{ss30}
\Gamma _{IMF}=\frac 16\int_{x_0}^1V_0(x)\left( \frac{x^2+x_0^2}{x^2}\right)
dx.
\end{equation}
Similar modification of Eq. (\ref{ss28}) gives 
\begin{equation}  \label{ss31}
\Gamma _{lab}=\frac 16\int_{x_0}^1V_0(x)\left( \frac{4x_0^2}{x^2+x_0^2}%
\right) dx.
\end{equation}
Obviously, both the integrals are equal for $V_0$ sharply peaked around $%
x=x_0$, but generally, for nonstatic quarks 
\begin{equation}  \label{ss32}
\Gamma _{lab}<\Gamma _{IMF}.
\end{equation}
Therefore, starting from the one test quark distribution function $V_0$ we
get the two different results on the first moment of the function $g_1$ -
depending on the used relation connecting distribution and structure
functions.

\begin{figure}[tbp]
\begin{center}
\epsfig{file=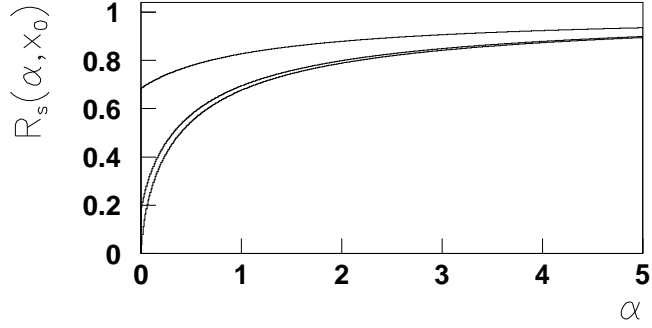, height=6cm}
\end{center}
\caption{Ratio $R_s $ plotted for values $x_0 $=0.2, 0.02, 0.0002 - in order
from top to bottom.}
\label{yy3}
\end{figure}

What can our result (\ref{ss32}) mean quantitatively? Obviously, it will
depend on the function $V_0$, on its width. To get some feeling, we use for
the $V_0$ the following parameterization. According to the Eq. (\ref{eq20})
for $x>x_0$ one can write 
\begin{equation}  \label{m1}
xV_0^{\prime }(x)=-\frac M2P(p_0),\qquad p_0=\frac M2\left( x+ \frac{x_0^2}%
x\right) .
\end{equation}
Now, for $p_0$ close to $m$ let us parameterize the energy distribution by 
\begin{equation}  \label{m2}
P(p_0)\approx \exp \left( -\alpha \frac{p_0}m\right)
\end{equation}
satisfying the normalization 
\begin{equation}  \label{m3}
\int_m^\infty P(p_0)dp_0=1.
\end{equation}
Obviously, the distribution (\ref{m2}) means the average quark kinetic
energy equals to $m/\alpha $. Inserting (\ref{m2}) into (\ref{m1}) gives 
\begin{equation}  \label{m4}
V_0^{\prime }(x)\approx -\exp \left( -\frac \alpha 2\left[ \frac x{x_0}+%
\frac{x_0}x\right] \right) .
\end{equation}
Let us note, for $\left| y\right| \ll 1$%
\begin{equation*}
(1+y)^a\approx \exp (ay),
\end{equation*}
therefore if we substitute the exponential function in (\ref{m4}) by 
\begin{equation}  \label{m4a}
\exp \left( -\frac \alpha 2\left[ \frac x{x_0}+\frac{x_0}x\right] \right)
\sim \left[ \left( 1-x\right) \left( 1-\frac{x_0^2}x\right) \right] ^{\alpha
/2x_0}\equiv f(x,x_0),
\end{equation}
\begin{equation*}
x_0^2\leq x\leq 1,
\end{equation*}
then the resulting $V_0(x)$ will coincide with (\ref{m4}) in a vicinity of $%
x_0$, i.e. for small kinetic energies, but moreover will obey the global
kinematical constraint (\ref{eq24a}) outlined in Fig. \ref{yy2}. The ratio
of integrals (\ref{ss31}) and (\ref{ss30}) calculated by parts with the use
of Eqs. (\ref{m4}) and (\ref{m4a}) gives 
\begin{equation}  \label{m5}
R_s(\alpha ,x_0)\equiv \frac{\Gamma _{lab}}{\Gamma _{IMF}}=\frac{%
4\int_{x_0}^1x_0/x\left( \arctan [x/x_0]-\pi /4\right) f(x,x_0)dx}{%
\int_{x_0}^1\left( 1-x_0^2/x^2\right) f(x,x_0)dx}.
\end{equation}
The results of the numerical computing are plotted in the Fig. \ref{yy3}. 
What do these curves mean? It is obvious, that for static quarks, for which $%
\alpha \rightarrow \infty $ and $R_s\rightarrow 1$ both the approaches are
equivalent, as we have already shown. On the other hand, it is also
apparent, that for nonstatic quarks, with small ratio $\alpha \approx
m/\left\langle E_{kin}\right\rangle $, both the approaches can differ
substantially.

\subsection{Covariant formulation}

\label{sec2}

Master equation (\ref{ss1}) is assembled for quark momenta distributions $%
G,H $ in the nucleon rest frame, but despite of that, the equation is
relativistically covariant. Its manifestly covariant form follows
immediately from Eq. (\ref{ss1}) after the substitution 
\begin{equation}  \label{ms22}
G(p_{0,lab})=G\left( \frac{pP}M\right) ,\qquad H(p_{0,lab})=H\left( \frac{pP}%
M\right) .
\end{equation}
For moving nucleon we have 
\begin{equation}  \label{ms23}
P=(P_0,P_1,0,0),\qquad \beta =\frac{P_1}{P_0},\qquad \gamma = \frac{P_0}M
\end{equation}
and 
\begin{equation}  \label{ms24}
\frac{pP}M=\frac{p_0P_0-p_1P_1}M=\gamma (p_0-\beta p_1)=p_{0,lab},
\end{equation}
which means, that the phase space of the subset of quarks with $p_{0,lab}$
fixed, represented by the sphere 
\begin{equation}  \label{ms25}
p_1^2+p_2^2+p_3^2=p_{0,lab}^2-m^2
\end{equation}
in the nucleon rest frame, is in a boosted system correctly represented by
the ellipsoid with the shape defined by the Lorentz transform (\ref{ms24}).
Let us remark, in the same way as the Eq. (\ref{ss1}) a similar equation can
be obtained and solved also for the set of the neutrino structure functions,
nevertheless in this paper we consider only the electromagnetic ones.

Further, one can obtain also covariant solution of Eq. (\ref{ss1}) for the
spin functions $G_{1},G_{2}$, but first it is necessary to define correct
form of the quark polarization vector $w$. Generally, this vector should be
constructed from the proton momentum $P$, proton polarization vector $S$ and
the quark momentum $p$: 
\begin{equation}
w_{\mu }=AP_{\mu }+BS_{\mu }+Cp_{\mu },  \label{cr1}
\end{equation}%
where $A,B,C$ are invariant functions of $P,S,p.$ Then contracting of Eq. (%
\ref{ss5}) with $P_{\sigma },q_{\sigma }$ and $S_{\sigma }$ gives the
equations 
\begin{equation}
-SqMG_{2}=\frac{m}{2M\nu }\int H\left( \frac{pP}{M}\right) \left(
AM^{2}+CpP\right) \delta \left( \frac{pq}{M\nu }-x\right) \frac{d^{3}p}{p_{0}%
},  \label{cr2}
\end{equation}%
\begin{equation}
SqMG_{1}=\frac{m}{2M\nu }\int H\left( \frac{pP}{M}\right) \left( AM\nu
+BSq+Cpq\right) \delta \left( \frac{pq}{M\nu }-x\right) \frac{d^{3}p}{p_{0}},
\label{cr3}
\end{equation}%
\begin{equation}
-MG_{1}-\nu G_{2}=\frac{m}{2M\nu }\int H\left( \frac{pP}{M}\right) \left(
-B+CpS\right) \delta \left( \frac{pq}{M\nu }-x\right) \frac{d^{3}p}{p_{0}}.
\label{cr4}
\end{equation}%
Elimination of $G_{1},G_{2}$ gives 
\begin{equation}
\int H\left( \frac{pP}{M}\right) Cp\left( \frac{\nu P/M-q}{Sq}-S\right)
\delta \left( \frac{pq}{M\nu }-x\right) \frac{d^{3}p}{p_{0}}=0  \label{cr5}
\end{equation}%
and since $P,q,S$ are independent, $C$ must be zero. The remaining
invariants $A,B$ follow from Eq. (\ref{ss4}), which implies 
\begin{equation}
A^{2}M^{2}-B^{2}=-1,\qquad ApP+BpS=0  \label{cr6}
\end{equation}%
and solution of these equations reads 
\begin{equation}
A=\mp \frac{pS}{\sqrt{(pP)^{2}-(pS)^{2}M^{2}}},\qquad B=\pm \frac{pP}{\sqrt{%
(pP)^{2}-(pS)^{2}M^{2}}}.  \label{cr7}
\end{equation}%
So the quark polarization vector has the form 
\begin{equation}
w_{\mu }=\pm \frac{(pP)S_{\mu }-(pS)P_{\mu }}{\sqrt{(pP)^{2}-(pS)^{2}M^{2}}}.
\label{cr8}
\end{equation}%
Contributions of both possible solutions [sign $+(-)$ means that quark spin
is parallel (antiparallel) to the proton spin in its rest frame] are in our
calculation taken into account by the difference in Eq. (\ref{ss3}).
Apparently, for the proton rest frame and polarization $S=(0,1,0,0)$ the
last equation is identical to Eq. (\ref{ss12}). Now, the obtained invariants 
$A,B,C$ give the spin structure functions from Eqs. (\ref{cr2}), (\ref{cr3})
in covariant form 
\begin{equation}
G_{1}=\frac{m}{2M(Pq)(Sq)}\int H\left( \frac{pP}{M}\right) \frac{%
(pP)(Sq)-(pS)(Pq)}{\sqrt{(pP)^{2}-(pS)^{2}M^{2}}}\delta \left( \frac{pq}{Pq}%
-x\right) \frac{d^{3}p}{p_{0}},  \label{cr9}
\end{equation}%
\begin{equation}
G_{2}=\frac{mM}{2(Pq)(Sq)}\int H\left( \frac{pP}{M}\right) \frac{pS}{\sqrt{%
(pP)^{2}-(pS)^{2}M^{2}}}\delta \left( \frac{pq}{Pq}-x\right) \frac{d^{3}p}{%
p_{0}}.  \label{cr10}
\end{equation}%
Apparently, according to these relations the structure functions \textit{can
depend also on mutual orientation of $S$ and $q.$} Of course, this
dependence is more complicated, apart the factor $Sq$ ahead of the
integrals, integration involves also the terms $pS$ and $pq$. This question
is being studied and will be discussed in a separate paper. Our further
considerations will be based on the results obtained in the previous
section, which follow from Eqs. (\ref{cr9}), (\ref{cr10}) applied in the
proton rest frame for the longitudinal polarization $S=(0,1,0,0)$.
Obviously, for this case the last two equations are equivalent to Eqs. (\ref%
{ss13}), (\ref{ss14}).

The scheme based on the Eqs. (\ref{ss1}) and (\ref{cr8}) with all their
implications suggested in the previous sections can be a priori valid for
quasifree quarks (on mass shell) filling up the nucleon volume. In this
sense the scheme represents a covariant formulation of the naive QPM. We
have shown that Eq. (\ref{ss1}) in which the quark internal motion is
consistently taken into account implies the relations between the structure
and distribution functions different from those obtained in the standard
procedure relying on the preferred reference frame, IMF, which is based on
the approximation $p_{\mu }=xP_{\mu }$. In the covariant formulation this
approximation is equivalent to the assumption, that the partons are static
with respect to the nucleon. Of course, this consequence is somewhat
obscured just in the IMF, where all the relative motion is frozen, since all
the processes run infinitely slowly - including the passing of the probing
lepton through the nucleon. Let us remark, the standard relations (e.g. $%
F_{2}=x\sum e_{i}^{2}q_{i}$) obtained in the naive QPM with static quarks
are currently applied even in the standard approach based on QCD improved
QPM, which is not a consistent procedure, since it means that correct
dynamics is combined with incorrect kinematics.

In this way we have shown, that the relations between the structure and
distribution functions can be, at least on the level of the naive QPM,
strongly modified (particularly for the polarized case) by the parton
internal motion. This result can be instructive by itself. Let us remark,
the impact of the quark intrinsic motion on the function $g_1(x)$ has been
discussed also in some other approaches \cite{bo1}-\cite{rit} and necessity
of the covariant formulation for the spin structure functions has been
pointed out in \cite{cfs}.

\section{Model}

\label{sec3}

\begin{figure}[tbp]
\begin{center}
\epsfig{file=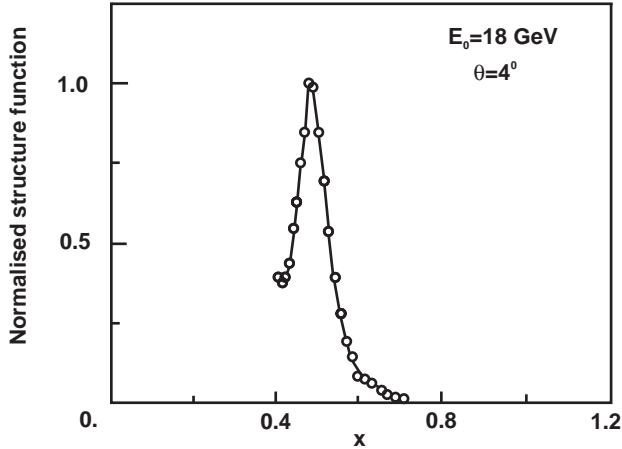, height=6cm}
\end{center}
\caption{The structure function for quasielastic $e^{-}d$ scattering, see
text }
\label{yy4}
\end{figure}

To illustrate, that the scheme suggested in the previous section can be
really valid for quasifree fermions, let us look at the Fig. \ref{yy4}, 
where the ''structure function'' of the deuteron measured in quasielastic $%
e^{-}d$ scattering \cite{D2} is shown, clearly proving the presence of two
nucleons in the nucleus. The similarity with the Fig. \ref{yy2} is apparent.

Of course, in the case of partons inside the nucleon the situation is much
more delicate. The interaction among the quarks and gluons is very strong,
partons themselves are mostly in some shortly living virtual states, is it
possible to speak about their mass at all? Strictly speaking probably not.
The mass in the exact sense is well defined only for free particles, whereas
the partons are never free by definition. However let us try to assume the
following. The relations obtained within the scheme suggested in the
previous sections can be used as a good approximation even for the
interacting quarks, but provided that the term \textit{mass of quasifree
parton} is substituted by the term \textit{parton effective mass}. By this
term we mean the mass, which a free parton would have to have to interact
with the probing photon equally as the real, bounded one. Intuitively, this
mass should correlate to $Q^{2}$: a lower $Q^{2}$ roughly means, that the
photon ''sees'' the quark surrounded by some cloud of gluons and
quark-antiquark pairs as a one particle - by which this photon is absorbed.
And on contrary, the higher $Q^{2}$ should mediate interaction with more
''isolated'' quark. Moreover, we accept that the value of the effective mass
can even for a fixed $Q^{2}$ fluctuate - e.g. in a dependence on the actual
QCD process accompanying the photon momentum transfer. This means, that the
terms in the relations involving the mass of quasifree parton $x_{0}=m/M$
will be substituted by their convolution with some 'mass distribution' $\mu
: $%
\begin{equation}
f(x,x_{0})\rightarrow \int \mu (x_{0},Q^{2})f(x,x_{0})dx_{0}=F(x,Q^{2}).
\label{cr11}
\end{equation}

In the following we shall propose a simple, but sufficiently general model
for the unknown distributions $\mu ,G,H,$ in which all the dynamics of the
system is absorbed. Then, these distributions will be used as an input for
the calculating of the corresponding structure functions. Construction of
the model is based on the following assumptions and considerations:

1) Parton distribution $P(\epsilon )d\epsilon $ representing the number of
quarks in the energy interval $\left\langle \epsilon ,\epsilon +d\epsilon
\right\rangle $ can be formally expressed: 
\begin{equation}  \label{mss31}
P(\epsilon )=\sum_jr_jj\rho _j(\epsilon ),\qquad \sum_jr_j=1,
\end{equation}
where $r_j$ is a probability that the nucleon is in the state with $j$
partons (quarks + antiquarks) of various flavors, and $\rho _j$ is the
corresponding average one-parton distribution, which satisfies 
\begin{equation}  \label{mss32}
\int \rho _j(\epsilon )d\epsilon =1.
\end{equation}

2) Nucleon consists of the three quarks and partons (gluons + $q\overline{q}$
pairs) mediating the interaction between them, as sketched in the Fig. \ref%
{yy5}\textit{a}, where the individual pictures represent some terms in the
sum (\ref{mss31}). 
\begin{figure}[tbp]
\begin{center}
\epsfig{file=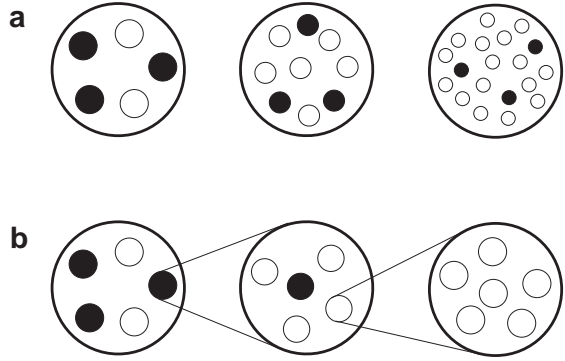, height=5cm}
\end{center}
\caption{Nucleon consisting of the valence and sea quarks - with different
resolutions, see text.}
\label{yy5}
\end{figure}
The flavors and spins of all the quarks in each the picture are mutually
cancelled, up to the three quarks giving additively the corresponding
nucleon quantum numbers. These three quarks are in the figure marked by
black and in our approach are identified with the \textit{valence quarks}.
The reason, that such identification is quite sensible, is the following.
Apparently, the sum (\ref{mss31}) can be split into quark and antiquark
parts $P_{q}(\epsilon ),P_{\overline{q}}(\epsilon )$, then our valence term
reads 
\begin{equation}
P_{val}(\epsilon )=P_{q}(\epsilon )-P_{\overline{q}}(\epsilon ),
\label{mssa32}
\end{equation}%
which in the $x-$representation exactly corresponds to the current
definition of the valence quarks. Correspondingly, the unmarked quarks are
identified with the \textit{sea quarks}. But both the kinds of quarks have
the same energy distributions $\rho _{j}(\epsilon )$ entering the Eq. (\ref%
{mss31}), in this sense they are completely equivalent. On the other hand,
it is obvious, that for the valence quarks, in Eq. (\ref{mss31}) only
''black'' quarks from the figure contribute, therefore if $\rho _{j}$ is in
the first approximation assumed to be independent on the flavor, then 
\begin{equation}
P_{val}(\epsilon )=3\sum_{j=3}^{\infty }r_{j}\rho _{j}(\epsilon ).
\label{mss33}
\end{equation}

3) The quarks carry only part of the nucleon energy (mass), 
\begin{equation}  \label{mss34}
\int P(\epsilon )\epsilon d\epsilon =c_qM,
\end{equation}
the rest is carried by the gluons. In the first approximation we shall
assume this factor is valid also for any term in the sum (\ref{mss31}), 
\begin{equation}  \label{mss35}
j\int \rho _j(\epsilon )\epsilon d\epsilon =c_qM,
\end{equation}
which in other words means the ratio of the total energies of quarks and
gluons, together constituting the nucleon mass, is the same for all possible
states sketched in Fig. \ref{yy5}\textit{a}.

4) We assume all the quarks in the nucleon state $j$ have approximately the
same effective mass $x_0=m_j/M$. One can expect, for higher $j$ the
parameter $x_0$ will drop and so the sum (\ref{mss33}) can be substituted by
the integral 
\begin{equation}  \label{mss37}
P_{val}(\epsilon )=3\int \mu _V(x_0)\rho (\epsilon ,x_0)dx_0,\qquad \int \mu
_V(x_0)dx_0=1.
\end{equation}
Obviously, Eq. (\ref{mss31}) can be with the use of Eq. (\ref{mss35})
rewritten in a similar way: 
\begin{equation}  \label{mss38}
P(\epsilon )=\int \mu (x_0)\rho (\epsilon ,x_0)dx_0,
\end{equation}
where 
\begin{equation}  \label{mssa38}
\mu (x_0)=\mu _V(x_0)\frac{c_qM}{\overline{\epsilon }(x_0)},\qquad \overline{%
\epsilon }(x_0)=\int \rho (\epsilon ,x_0)\epsilon d\epsilon .
\end{equation}
The physical meaning of the distributions $\mu _V,\mu $ is the following.
The distribution $\mu (x_0)$ represents a probability, that the effective
mass of the quark, which the probing lepton interacts with, is $x_0$ or
alternatively, $\mu (x_0)dx_0$ is the number of quarks in the interval $%
\left\langle x_0,x_0+dx_0\right\rangle $, which the lepton has chance to
interact with. On the other hand, the (normalized) distribution $\mu _V(x_0)$
can be interpreted as a probability, that the exchanging photon
''distinguishes'' the quarks with the effective mass $x_0$ - as expressed by
the pictures with different granularity in Fig. \ref{yy5}\textit{a.} In this
sense, each picture in the Fig. \ref{yy5}\textit{a} can be labeled by some $%
x_0$, equally as the corresponding term $\rho (\epsilon ,x_0)$ in the
integral (\ref{mss37}). Obviously, at the same time the $\mu _V(x_0)$
represents also the distribution of effective masses corresponding to the
valence quark term. Intuitively, the probability of different contributions
in Fig. \ref{yy5}\textit{a} should depend also on $Q^2$ (higher $Q^2=$%
'better resolution'), so we expect

\begin{equation}  \label{mss39}
\mu (x_0)\rightarrow \mu (x_0,Q^2),\qquad \mu _V(x_0)\rightarrow \mu
_V(x_0,Q^2).
\end{equation}
In the next, we shall identify these distributions with that introduced in
Eq. (\ref{cr11}).

5) The relations (\ref{r9}), (\ref{r12}) and (\ref{eq20}) give the recipe
how to obtain the structure function $F_2$ from a given energy distribution
of the partons with some fixed value $x_0$ and charge $e_q:$%
\begin{equation*}
F_2(x,x_0)=e_q^2\varphi (x,x_0),\qquad \varphi (x,x_0)=-x^2\int_x^1\frac{%
2V_0^{\prime }(\xi )\xi }{\xi ^2+x_0^2}d\xi ,
\end{equation*}
\begin{equation}  \label{mss310}
2V_0^{\prime }(\xi )\xi =\pm M\rho (\epsilon ,x_0),\qquad \epsilon =\frac
M2\left( \xi +\frac{x_0^2}\xi \right) ,
\end{equation}
where the sign $+(-)$ in the second relation refers to the region $\xi <x_0\
(\xi >x_0).$ For the application of this procedure to Eqs. (\ref{mss37}), (%
\ref{mss38}) one has to weight the contributions integrated over $x_0$ by
the corresponding (mean) charge squared. Apparently, the charge weight of
the valence quarks is the constant 
\begin{equation}  \label{mss311}
w_{val}=\frac 13\left[ \left( \frac 23\right) ^2+\left( \frac 23\right)
^2+\left( \frac 13\right) ^2\right] =\frac 13
\end{equation}
for the proton and similarly for the neutron, $w_{val}=2/9.$ For the sea we
assume in the first approximation the ''equilibrated mixture'' of the quarks 
$u:d:s=1:1:1$, so 
\begin{equation}  \label{mss312}
w_{sea}=\frac 13\left[ \left( \frac 23\right) ^2+\left( \frac 13\right)
^2+\left( \frac 13\right) ^2\right] =\frac 29.
\end{equation}
Then for the nucleon with $j$ quarks we get 
\begin{equation}  \label{mss313}
w_j=\frac{3w_{val}+(j-3)w_{sea}}j=w_{sea}+\frac{3(w_{val}-w_{sea})}j,
\end{equation}
or in terms of $x_0$ 
\begin{equation}  \label{mss314}
w(x_0)=w_{sea}+3(w_{val}-w_{sea})\frac{\overline{\epsilon }(x_0)}{c_qM}.
\end{equation}
Therefore, the energy distributions (\ref{mss37}), (\ref{mss38}) generate
the corresponding structure functions: 
\begin{equation}  \label{mss315}
F_{2val}(x,Q^2)=3w_{val}\int \mu _V(x_0,Q^2)\varphi (x,x_0)dx_0,
\end{equation}
\begin{equation}  \label{mss316}
F_2(x,Q^2)=\int \left( \frac{c_qM}{\overline{\epsilon }(x_0)}%
w_{sea}+3(w_{val}-w_{sea})\right) \mu _V(x_0,Q^2)\varphi (x,x_0)dx_0.
\end{equation}

6) Now, let us pay attention to the spin structure functions. According to
the concept suggested in the item 2), only valence quarks contribute to the
nucleon spin. First, we shall consider the spin functions generated by the
valence quarks with some fixed effective mass $x_0,$ then we shall easily
proceed to the case with the distribution $\mu _V(x_0)$.

In Sec. \ref{spin} we have simply assumed all the three valence quarks
contribute to the proton spin equally [Eq. (\ref{ss7})]. On the other hand
it is obvious the quark symmetry group can impose an extra constraint on the
contributions of different quark flavors as it follows e.g. from the
philosophy of the well known Bjorken \cite{bjo} and Ellis-Jaffe \cite{eja}
sum rules based on the symmetries \textit{U(6)} and \textit{SU(3)}. If we do
not assume any particular group of symmetry, then the different spin
contributions of \textit{u-} and \textit{d-}quarks can be expressed by the
free parameter $a,\ 0\leq a\leq 1$, having in the notation of Eq. (\ref{ss7}%
), e.g. for the proton, the following sense 
\begin{equation}
\Delta h_{u}(p_{0})=2ah(p_{0}),\qquad \Delta h_{d}(p_{0})=(1-2a)h(p_{0}),
\label{ms31}
\end{equation}%
where $h$ is the valence distribution 
\begin{equation}
u(p_{0})=d(p_{0})\equiv h(p_{0}),\qquad \int h(p_{0})d^{3}p=1,  \label{ms31a}
\end{equation}%
which is not, due to different normalization, identical with the
distribution $\rho (\epsilon )$, but the both are simply related 
\begin{equation}
\rho (\epsilon )=4\pi \epsilon \sqrt{\epsilon ^{2}-m^{2}}h(\epsilon ),
\label{ms32}
\end{equation}%
in the same way, as the distributions $P,G$ in Eq. (\ref{eq18}).

In the case of proton, there are the particular cases:

\noindent a) $a=0$ corresponds to the mutual spin orientation of the three
valence quarks $(s_u,s_u,s_d)=(-1,+1,+1).$

\noindent b) $a=1/3$ corresponds to the oversimplified scenario studied in
Sec. \ref{spin}, assuming the equal contribution of all the three quarks; $%
(s_u,s_u,s_d)=(+1/3,+1/3,+1/3).$

\noindent c) $a=2/3$ corresponds to the non-relativistic \textit{SU(6)}
approach. From the wave function%
\begin{equation*}
\mid p,\uparrow \rangle =\frac 1{\sqrt{2}}\left( \frac 1{\sqrt{6}}\left|
(ud+du)u-2uud\right\rangle \otimes \frac 1{\sqrt{6}}\left| (\uparrow
\downarrow +\downarrow \uparrow )\uparrow -2\uparrow \uparrow \downarrow
\right\rangle \right.
\end{equation*}
\begin{equation}  \label{ms32a}
+\frac 1{\sqrt{2}}\left. \left| (ud-du)u\right\rangle \otimes \frac 1{\sqrt{2%
}}\left| (\uparrow \downarrow -\downarrow \uparrow )\uparrow \right\rangle
\right)
\end{equation}
one can easily show the mean value of the spin carried by the $d(u)-$ quarks
is $-1/3(4/3)$, i.e. $(s_u,s_u,s_d)=(+2/3,+2/3,-1/3)$, which agrees with $%
a=2/3$ in Eq. (\ref{ms31}).

\noindent d) $a=1$ corresponds to the mutual orientation of the three quarks 
$(s_u,s_u,s_d)=(+1,+1,-1).$

So, the proton spin function $H(p_{0})$ entering the equations (\ref{ss17}),
(\ref{ss18}) and expressed in terms of the functions (\ref{ms31}) reads 
\begin{equation}
H^{p}(p_{0})=\frac{8}{9}au(p_{0})+\frac{1}{9}(1-2a)d(p_{0})=\frac{1}{9}%
(1+6a)h(p_{0}).  \label{ms33}
\end{equation}%
Assuming the neutron is isospin symmetric, its corresponding spin function
will be 
\begin{equation}
H^{n}(p_{0})=\frac{4}{9}(1-2a)u(p_{0})+\frac{2}{9}ad(p_{0})=\frac{1}{9}%
(4-6a)h(p_{0}),  \label{ms34}
\end{equation}%
therefore the corresponding equations for the nucleon spin structure
functions read 
\begin{equation*}
g_{j}^{p}(x,x_{0})=w_{spin}^{p}\psi _{j}(x,x_{0}),\qquad
g_{j}^{n}(x,x_{0})=w_{spin}^{n}\psi _{j}(x,x_{0}),\qquad j=1,2,
\end{equation*}%
\begin{equation}
w_{spin}^{p}=\frac{1}{9}(1+6a),\qquad w_{spin}^{n}=\frac{1}{9}(4-6a),
\label{ms35}
\end{equation}%
where, in accordance with Eqs. (\ref{ss17}), (\ref{ss18}) 
\begin{equation}
\psi _{1}(x,x_{0})=\frac{m}{2}\int \frac{h(p_{0})}{p_{0}}\sqrt{\frac{%
p_{0}+p_{1}}{p_{0}-p_{1}}}\delta \left( \frac{p_{0}+p_{1}}{M}-x\right)
d^{3}p,  \label{ms37}
\end{equation}%
\begin{equation}
\psi _{2}(x,x_{0})=-\frac{m}{2}\int \frac{h(p_{0})}{p_{0}}\frac{p_{1}}{\sqrt{%
p_{0}^{2}-p_{1}^{2}}}\delta \left( \frac{p_{0}+p_{1}}{M}-x\right) d^{3}p.
\label{ms38}
\end{equation}%
The function $\psi _{1}(x,x_{0})$ can be with the use of Eqs. (\ref{ss23}), (%
\ref{ss26}) expanded 
\begin{equation}
\psi _{1}(x,x_{0})=\frac{x_{0}}{2}\sum_{j=0}^{\infty }{\binom{{-\frac{1}{2}}%
}{j}}(-1)^{j}V_{-j-3/2}(x)\left( \frac{x}{2}\right) ^{j+1/2}.  \label{ms324}
\end{equation}%
%
%
%
%
%
Since Eq. (\ref{r12}) implies 
\begin{equation}
V_{-j-3/2}(x)=-\int_{x}^{1}V_{0}^{\prime }(\xi )\left( \frac{2\xi }{\xi
^{2}+x_{0}^{2}}\right) ^{j+3/2}d\xi ,  \label{ms325}
\end{equation}%
one can easily show the sum in Eq. (\ref{ms324}) gives 
\begin{equation}
\psi _{1}(x,x_{0})=-x_{0}\int_{x}^{1}\frac{V_{0}^{\prime }(\xi )\xi }{\xi
^{2}+x_{0}^{2}}\sqrt{\frac{x\xi }{\xi ^{2}+x_{0}^{2}-x\xi }}d\xi .
\label{ms326}
\end{equation}%
Similar manipulation with the function $\psi _{2}$ gives the result 
\begin{equation}
\psi _{2}(x,x_{0})=-\frac{x_{0}}{2}\int_{x}^{1}\frac{V_{0}^{\prime }(\xi
)\xi }{\xi ^{2}+x_{0}^{2}}\frac{\xi ^{2}+x_{0}^{2}-2x\xi }{\sqrt{x\xi (\xi
^{2}+x_{0}^{2}-x\xi )}}d\xi .  \label{ms327}
\end{equation}%
Obviously, for the case with the distribution $\mu _{V},$ the corresponding
spin structure functions read 
\begin{equation}
g_{j}(x,Q^{2})=w_{spin}\int \mu _{V}(x_{0},Q^{2})\psi
_{j}(x,x_{0})dx_{0},\qquad j=1,2.  \label{ms328}
\end{equation}%
Let us note, the structure functions $F_{2},F_{2val},g_{1},g_{2}$ are not
independent, all of them are in the corresponding way generated by the
distributions $\mu _{V}$ and $V_{0}$ (or, equivalently by $\rho $).

7) Now, to make the construction suggested above applicable for a
quantitative comparison with the experimental data, we have to use some
reasonable, simple and sufficiently flexible parameterization for the
unknown functions $\mu _V$ and $V_0$. We suggest the following.

\noindent a)\textit{\ }Normalized distribution $\mu _V$ is assumed in the
form 
\begin{equation}  \label{ms329}
\mu _V(x_0,Q^2)=\frac{\Gamma (r+s+2)}{\Gamma (r+1)\Gamma (s+1)}\cdot
x_0^r(1-x_0)^s,\qquad 0<x_0<1,
\end{equation}
where the $Q^2-$dependence is involved in the parameters $r,s.$

\noindent b)\textit{\ }For the function $V_{0}^{\prime }(x)x$ we shall use
the parameterization suggested in Eqs. (\ref{m2})-(\ref{m4a}) 
\begin{equation}
V_{0}^{\prime }(x)x=\mp c_{norm}\cdot f(x,x_{0}),\qquad f(x,x_{0})\equiv 
\left[ (1-x)\left( 1-\frac{x_{0}^{2}}{x}\right) \right] ^{\alpha /2x_{0}},
\label{ms330}
\end{equation}%
where the upper (lower) sign in the first relation refers to the region $%
x>x_{0}\quad (x<x_{0})$ and 
\begin{equation}
c_{norm}=\left[ \int_{x_{0}}^{1}f(x,x_{0})\left( 1-\frac{x_{0}^{2}}{x^{2}}%
\right) dx\right] ^{-1},  \label{ms331}
\end{equation}%
which follows e.g. from Eqs. (\ref{m3}), (\ref{eq25}). Now, apparently one
has to accept the parameter $\alpha \approx m/\left\langle
E_{kin}\right\rangle $ depends on $x_{0}$ as well. Let us consider the
following. Sequence of the pictures in Fig. \ref{yy5}\textit{a} can be
understood as the pictures of the one and the same nucleon, but ''seen with
different resolutions'' as outlined in Fig. \ref{yy5}\textit{b}. Then, it is
natural to assume the momentum $P$ of the parton from some picture can be
obtained from the momenta $p_{\lambda }$ of $n$ partons in a picture more
rightward, representing the parton ''seen with better resolution'': 
\begin{equation}
P=\sum_{\lambda }p_{\lambda }.  \label{ms332}
\end{equation}%
Obviously, the mean values satisfy 
\begin{equation}
\left\langle P_{0}\right\rangle =n\left\langle p_{\lambda 0}\right\rangle
,\qquad \left\langle \left| \vec{P}\right| \right\rangle =c_{corr}\cdot
n\left\langle \left| \vec{p}_{\lambda }\right| \right\rangle ,\qquad 0\leq
c_{corr}\leq 1,  \label{ms333}
\end{equation}%
where $c_{corr}$ equals $0(1)$ for the extreme case, when the motion of the
partons in the corresponding subset is completely uncorrelated (correlated).
The last relations imply the effective masses and kinetic energies obey 
\begin{equation}
m(P)\geq n\cdot m(p),\qquad \left\langle E_{kin}(P)\right\rangle \leq n\cdot
\left\langle E_{kin}(p)\right\rangle ,  \label{ms334}
\end{equation}%
which means the quantity $\alpha $ is a non-decreasing function of $x_{0}$.
In this moment we know nothing more about this function, in the next section
we shall show, that a reasonable agreement with the experimental data can be
obtained with the parameterization: 
\begin{equation}
\alpha (x_{0})=\alpha _{1}\left( -\ln (x_{0})\right) ^{-\alpha _{2}}.
\label{ms335}
\end{equation}

Since we parameterize the function $V_0^{\prime }$ rather than the function $%
\rho $, it will be useful to express the quantity $\overline{\epsilon }%
(x_0), $ defined in Eq. (\ref{mss38}) and afterwards entering the important
Eq. (\ref{mss316}), also in terms of $V_0^{\prime }$. Obviously, using Eqs. (%
\ref{mss310}) and (\ref{ms330}) one gets 
\begin{equation}  \label{ms336}
\overline{\epsilon }(x_0)=\int \rho (\epsilon ,x_0)\epsilon d\epsilon
=-\int_{x_0}^1V_0^{\prime }(\xi )\xi \left( \xi +\frac{x_0^2}\xi \right)
\frac M2\left( 1-\frac{x_0^2}{\xi ^2}\right) d\xi
\end{equation}
\begin{equation*}
=c_{norm}\frac M2\int_{x_0}^1f(\xi ,x_0)\left( \xi -\frac{x_0^4}{\xi ^3}%
\right) d\xi .
\end{equation*}

Now, we can our results shortly summarize. If there are given some values of
the free parameters $c_q,a,r,s,\alpha _1,\alpha _2$, then the corresponding
proton and neutron structure functions can be directly calculated according
to Eqs. (\ref{mss315}), (\ref{mss316}), (\ref{ms328}), where the
distribution $\mu _V$ is given by Eq. (\ref{ms329}), the function $\overline{%
\epsilon }(x_0)$ by Eq. (\ref{ms336}) with the use of Eqs. (\ref{ms330}), (%
\ref{ms331}), (\ref{ms335}) and the functions $\varphi ,\psi _1,\psi _2$
are: 
\begin{equation}  \label{ms337}
\varphi (x,x_0)=2x^2\int_x^1\eta (\xi ,x_0)d\xi ,
\end{equation}
\begin{equation}  \label{ms338}
\psi _1(x,x_0)=x_0\int_x^1\eta (\xi ,x_0)\sqrt{\frac{x\xi }{\xi
^2+x_0^2-x\xi }}d\xi ,
\end{equation}
\begin{equation}  \label{ms339}
\psi _2(x,x_0)=\frac{x_0}2\int_x^1\eta (\xi ,x_0)\frac{\xi ^2+x_0^2-2x\xi }{%
\sqrt{x\xi (\xi ^2+x_0^2-x\xi )}}d\xi ,
\end{equation}
where 
\begin{equation}  \label{ms340}
\eta (\xi ,x_0)=c_{norm}\theta (\xi -x_0)\frac{f(\xi ,x_0)}{\xi ^2+x_0^2}.
\end{equation}
The last expression is calculated from Eqs. (\ref{ms330}), (\ref{ms331}) and
(\ref{ms335}) with the use of the step function $\theta (x)=+1(-1)$ for $%
x>0\ (x<0).$

\section{Comparison with the experimental data}

\label{sec4}

Now we shall try to compare our formulas for the structure functions with
the existing data. We shall not attempt to make a consistent, global fit of
the free parameters based on some rigorous fitting procedure, but only show
the set of optimal parameters obtained by their tentative varying on the
computer ''by hand''. Moreover, our constraint will be only agreement with
the proton structure functions $F_2$ and $g_1$. It means that the parameter $%
a,$ controlling asymmetry between the proton and neutron spin functions,
must be somehow fixed before the fitting. For the first approximation we use
the $SU(6)$ value, $a=2/3$ [see item 6c) in the previous section].

\begin{figure}[tbp]
\begin{center}
\epsfig{file=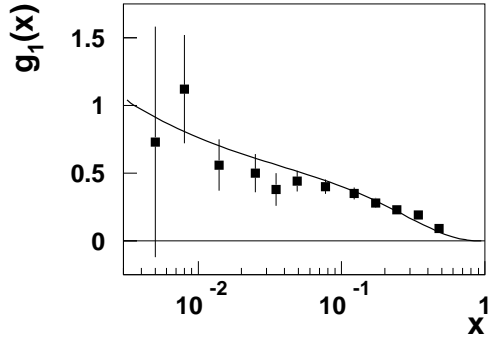, height=6cm}
\end{center}
\caption{Proton spin structure function $g_{1}(x)$ at $Q^{2}=10GeV^{2}/c^{2}$%
. The points represent experimental data {\protect\cite{smc}}, the curve is
the result of our calculation.}
\label{yy6}
\end{figure}

\begin{figure}[tbp]
\begin{center}
\epsfig{file=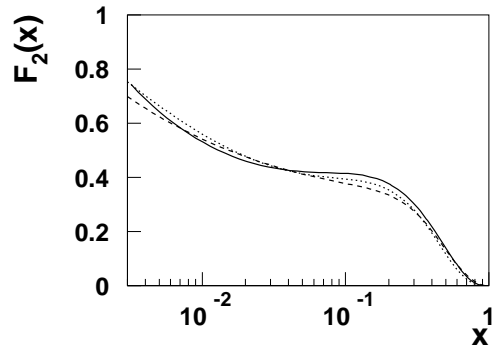, height=6cm}
\end{center}
\caption{Proton structure function $F_{2}(x)$ at $Q^{2}=10GeV^{2}/c^{2}$.
The dotted and dashed curves represent the fits of the experimental data
suggested in {\protect\cite{f2h1}} and {\protect\cite{smc}}. The full curve
is the result of our calculation.}
\label{yy7}
\end{figure}

For a comparison with $F_{2}$ we use the parameterizations of the world data
suggested in \cite{f2h1} and \cite{smc}, both taken for $%
Q^{2}=10GeV^{2}/c^{2}$. The data for $g_{1}$ are taken over from the paper %
\cite{smc} of the SMC Collaboration. After some checking on the computer,
the optimal set of the free parameters is considered: 
\begin{equation}
c_{q}=0.43,\quad r=-0.49,\quad s=6.5,\quad \alpha _{1}=1.6,\quad \alpha
_{2}=1.5  \label{ms41}
\end{equation}%
Results of the calculation of the proton structure functions $g_{1}$ and $%
F_{2}$ with these parameters are shown in Figs. \ref{yy6}, \ref{yy7}
together with the data. 
Let us remark, the experimental points for $g_{1}$ correspond to the values
evolved in \cite{smc} to $Q^{2}=10GeV^{2}/c^{2}$. In the error bars all the
quoted errors (statistical, systematic and those due to uncertainty of QCD
evolution) are combined. Obviously, the agreement with the experimental data
in both the figures can be considered very good, particularly if we take
into account that our parameterization of the unknown distributions is
perhaps the simplest possible and moreover, the parameters (\ref{ms41})
still may not be optimal.

\begin{figure}[tbp]
\begin{center}
\epsfig{file=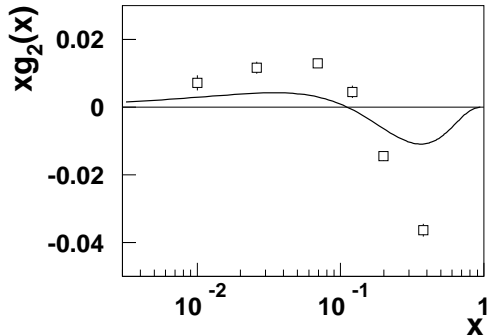, height=6cm}
\end{center}
\caption{Proton structure function $xg_{2}(x)$ at $Q^{2}=10GeV^{2}/c^{2}$.
The points represent the term $xg_{2}^{WW}(x)$ (see text), the curve is the
result of our calculation.}
\label{yy8}
\end{figure}

\begin{figure}[tbp]
\begin{center}
\epsfig{file=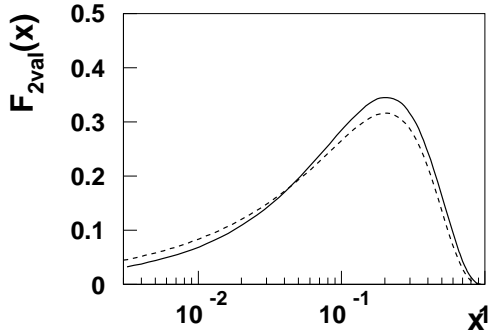, height=6cm}
\end{center}
\caption{Proton structure function $F_{2val}(x)$. The dashed curve
represents the function based on the standard global analysis according to
the relations (\ref{ms42}). The full curve is the result of our calculation.}
\label{yy9}
\end{figure}

Now, having ''tuned'' the free parameters by the $g_{1}$ and $F_{2}$, one
can predict the remaining functions $g_{2}$ and $F_{2val}$. The results are
shown in Figs. \ref{yy8}, \ref{yy9}. 
Our $xg_{2}$ surely does not contradict the experimental data \cite{smcd},
which are compatible with zero - with statistical errors bigger, than the
vertical range of the figure. Thus instead of the data, the comparison is
done with Wandzura Wilczek \cite{ww} twist-2 term for $xg_{2}^{WW}$, which
is evaluated in \cite{smcd} from the corresponding $g_{1}$. It is obvious,
that two completely different approaches give at least qualitatively very
similar results. The proton valence function $F_{2val}$ in Fig. \ref{yy9} is
compared with the corresponding combination of the distributions $xu_{V}(x)$
and $xd_{V}(x)$ obtained (for $Q^{2}=4GeV^{2}/c^{2}$) in \cite{msr} by the
standard global analysis: 
\begin{equation}
F_{2val}(x)=\frac{8}{9}xu_{V}(x)+\frac{1}{9}xd_{V}(x),\quad \int
u_{V}(x)dx=\int d_{V}(x)dx=1.\   \label{ms42}
\end{equation}%
Apparently, the agreement can be considered good. One can note, that the two
different procedures, the standard one (uses input on $F_{2},F_{3}^{\nu N}$
+ QCD) and ours (uses input on $F_{2},g_{1}$ + our model) give a very
similar picture of the function $F_{2val}(x)$, which is not directly
measurable.

\section{Discussion}

\label{sec5}

\begin{figure}[tbp]
\begin{center}
\epsfig{file=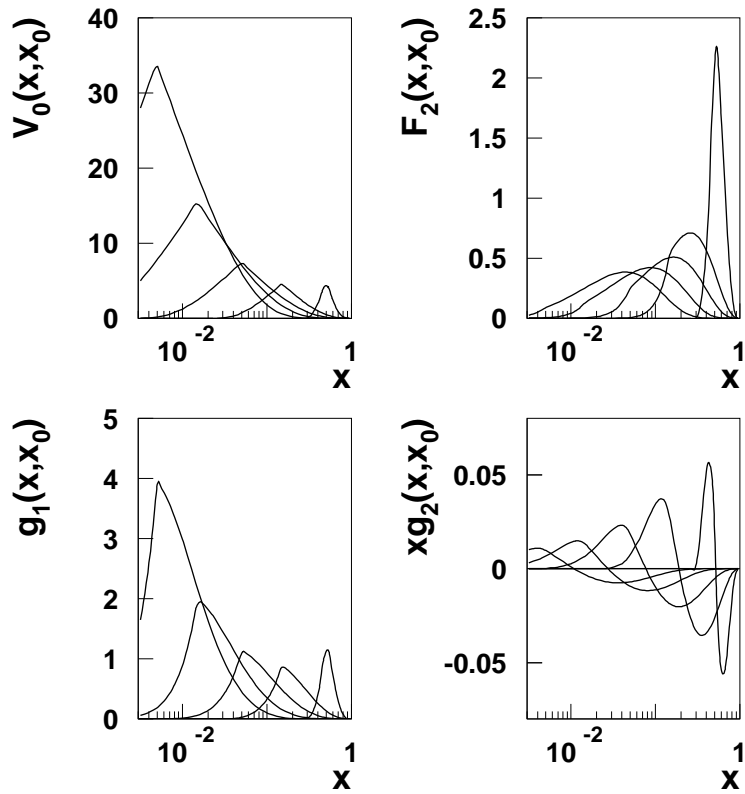, height=12cm}
\end{center}
\caption{Distribution functions $V_{0}(x,x_{0})$ drawn for $%
x_{0}=0.005,0.015,0.05,0.15,0.5$ and the structure functions generated
correspondingly. The calculation is based on the Eqs. (\ref{ms330}), (\ref%
{mss310}), (\ref{ms326}) and (\ref{ms327}).}
\label{yy10}
\end{figure}

Let us make a few comments on the obtained results. First of all, it should
be pointed out, that our structure functions in Figs. \ref{yy6}-\ref{yy9}
are calculated on the basis of very simple parameterization of the unknown
distributions $\mu (x_{0})$ and $V_{0}(x,x_{0})$, but on the other hand it
is essential, that the contributions from the individual components of the
quark distribution correctly take into account the intrinsic quark motion,
which is particularly important for the spin structure function. The effect
of this motion on $g_{1}$ is demonstrated in Fig. \ref{yy3} and the fact,
that we succeeded to achieve a good agreement with the data in Fig. \ref{yy6}
is just thanks to this effect. For a better insight, how our structure
functions are generated, in the Fig. \ref{yy10} we have displayed the
initial distribution function $V_{0}(x,x_{0})$ drawn for a few values $%
x_{0}, $ together with the corresponding structure functions $%
F_{2},g_{1},xg_{2}$. 
The complete structure functions are their superpositions - weighted by the
corresponding way with the use of the distribution $\mu _{V}(x_{0})$.

Further, also some other assumptions of the model are possibly
oversimplified, for a more precise calculation, at least some of them could
be rightly modified - but at a price of introducing the additional free
parameters. For example, the constant $w_{sea}$ [see Eq. (\ref{mss312})]
should take into account some suppression of the $s-$ quarks \cite{msr} and
probably should allow a weak dependence on $x_{0}$. Also for the constant $%
c_{q}$ [see Eq. (\ref{mss34})] some $x_{0}-$dependence should be allowed.
Concerning this constant, let us make one more comment. The standard global
fit \cite{msr} suggests (at $Q^{2}=10GeV^{2}/c^{2}$) the quarks carry $%
\simeq 56\%$ of the nucleon energy and our fitted value $c_{q}$ from the Eq.
(\ref{mss34}) is roughly $43\%$. This difference is mainly due to the
different relations between the distribution and structure function in both
the approaches, see Eqs. (\ref{sf4}), (\ref{r13}). The second relation
(valid for a subset of quarks with effective mass $x_{0}$), multiplied by $%
x^{2}$ and then integrated by parts gives 
\begin{equation}
\int_{x_{0}^{2}}^{1}xF(x,x_{0})dx=\frac{1}{4}%
\int_{x_{0}^{2}}^{1}F_{2}(x,x_{0})\left( 3+\frac{x_{0}^{2}}{x^{2}}\right) dx,
\label{ms51}
\end{equation}%
which for the static quarks [$F(x,x_{0})\simeq F_{2}(x,x_{0})/x\simeq \delta
(x-x_{0})$, see discussion after Eq. (\ref{r13})] coincides with the
standard relation. Nevertheless, generally both the relations imply
different rate of the nucleon energy carried by quarks. One can check
numerically that for our $F_{2}(x,x_{0})$ in a dominant region of $x_{0}$
the term $(x/x_{0})^{2}$ in the integral (\ref{ms51}) plays a minor role
(see also Fig. \ref{yy10}, positions of the maxima of $F_{2}$'s are above
the corresponding $x_{0}$, in particular for lower $x_{0}$), so as a result
we get $3/4$ of the standard estimation of the quark contribution to the
nucleon energy. This ratio agrees with the ratio obtained from the
corresponding fits: $3/4\simeq 43\%\ /\ 56\%.$

In the previous section we have assumed the same shape for the valence terms
related to the $u$ and $d-$quarks. This assumption together with our premise 
$a=2/3$ (i.e. $SU(6)$ symmetry) in an accordance with Eq. (\ref{ms34}) give $%
H^{n}(p_{0})=0$ and $g_{1}^{n}(x)=0$ correspondingly. On the other hand it
is known, that the neutron structure function $g_{1}^{n}(x)\neq 0$, even if $%
\left| \Gamma _{1}^{n}\right| $ is substantially less than $\Gamma _{1}^{p}$%
. A more consistent approach ($a=2/3$ but $d\neq u$) would give%
\begin{equation}
H^{n}(p_{0})=\frac{4}{27}(-u^{n}(p_{0})+d^{n}(p_{0}))=\frac{4}{27}%
(-d^{p}(p_{0})+u^{p}(p_{0})),  \label{msb51}
\end{equation}%
therefore $d\neq u$ implies $g_{1}^{n}(x)\neq 0$. Actually, the global fit
analysis proves that $d_{val}^{p}(x)$ is slightly ''narrower'' than $%
u_{val}^{p}(x)$. It means, considering qualitatively, in accordance with the
equation above in the function $g_{1}^{n}(x)$ the negative term should
dominate for smaller $x$, which does not contradict the data. A proper
accounting for this difference into the model should enable to calculate
consistently in a better approximation not only the proton and neutron
structure functions $F_{2},g_{1},g_{2}$, but also the neutrino structure
functions. Apparently then one could make a ''super-global'' fit covering
the both unpolarized and polarized DIS data. As a result, the
flavor-dependent quark distributions $V_{0}(x,x_{0})$ [or equivalently $\rho
(\epsilon ,x_{0})$] together with the corresponding effective mass
distributions and the parameter $a$ controlling the relative spin
contribution of the $u-$ and $d-$quarks, could be obtained.

Finally, let us point out, inclusion the spin structure function into the
fit in our model enables to obtain some information about the distribution
of the quark effective masses. Within our approach there are two
distributions, $\mu _{V}$ and $\mu $, relevant for the description of the
quark effective masses in the nucleon. The extrapolation of our
parameterization for the $\mu $ distribution with the use of the relations $%
\alpha \approx m/\left\langle E_{kin}\right\rangle $ and (\ref{mssa38}),(\ref%
{ms329}),(\ref{ms335}),(\ref{ms41}) give for $x_{0}\rightarrow 0:$%
\begin{equation}
\mu (x_{0})\sim \frac{\mu (x_{0})}{\overline{\epsilon }(x_{0})}\rightarrow 
\frac{x_{0}^{r}}{Mx_{0}+\left\langle E_{kin}\right\rangle }\rightarrow \frac{%
x_{0}^{-1.49}}{\left| \ln x_{0}\right| ^{1.5}},  \label{msa51}
\end{equation}%
which implies the extrapolated $\mu $ is not integrable in this limit. On
the other hand, the basic distribution $\mu _{V}$, parameterized by Eq. (\ref%
{ms329}) with the $r,s$ from the set (\ref{ms41}) and with the use of the
known relation $z\Gamma (z)=\Gamma (z+1)$ can give an estimate of the mean
value: 
\begin{equation}
\left\langle x_{0}\right\rangle _{V}=\frac{r+1}{r+s+2}\simeq 0.064,
\label{ms52}
\end{equation}%
i.e. $\left\langle m\right\rangle \simeq 60MeV$ for $Q^{2}=10GeV^{2}/c^{2}$.
The corresponding kinetic term calculated as 
\begin{equation}
\left\langle E_{kin}\right\rangle _{V}=\int \mu _{V}(x_{0})(\overline{%
\epsilon }(x_{0})-Mx_{0})dx_{0}  \label{msa52}
\end{equation}%
gives a similar number ($\simeq 60MeV$). This number agree very well with
the corresponding temperature obtained in the statistical model \cite{bha}.
The $Q^{2}$-dependence is involved only in the distribution $\mu
_{V}(x_{0},Q^{2})$, i.e. in our parameterization (\ref{ms329}) only via the
powers $r(Q^{2}),s(Q^{2})$. It follows, the structure functions, which
enhance in a low$-x$ region for increasing $Q^{2}$, must be generated by the
distribution $\mu _{V}(x_{0},Q^{2})$ in which the mean effective mass $%
\left\langle x_{0}\right\rangle _{V}$ drops for increasing $Q^{2}$ - in the
qualitative agreement with an intuitive expectation.

\section{Summary}

\label{sec6}

In the present paper we proposed an alternative, covariant formulation of
the QPM. The initial postulates of the standard and our approach are
basically the same, despite of that the relations between the structure and
distribution functions obtained in both the approaches are not identical. It
is due to the fact, that in the standard approach the intrinsic quark motion
is effectively suppressed by the use of the approximation $p_{\mu }=xP_{\mu
} $. On the other hand, we have shown the master equations can be solved
without the use of the this approximation, so in the corresponding solution
the quark intrinsic motion is consistently taken into account. In
particular, we have suggested, that the quark intrinsic motion can
substantially reduce the structure function $g_1$.

On the basis of the obtained relations (a priori valid for the version of
naive covariant QPM - with nonstatic quarks on mass shell) we propose the
model, in which the distributions $(\mu _{V},V_{0})$ reflecting the parton
dynamics are introduced with some free parameters. With the use of this
model we calculated simultaneously the proton structure functions $%
F_{2},F_{2val},g_{1},g_{2}$, assuming only the valence quarks term
contributes to the proton spin. Then by a comparison with the data ($%
F_{2},g_{1};Q^{2}=10GeV^{2}/c^{2}$) we fixed the free parameters. We found
out:

1) Both the unpolarized structure functions can be well reproduced by the
model. The comparison is done with the data on $F_{2}$ and with the $%
F_{2val} $ obtained from the standard global analysis data.

2) At the same time, the model well agrees with the data on $g_1$ and the
calculated $g_2$ does not contradict the existing experimental data.

3) Analysis of the fixed parameters within our approach suggests:

\noindent $i)$ The quarks carry less the proton energy (almost by the factor
3/4), than estimated from the standard analysis.

\noindent $ii)$ The average effective mass related to the valence quark term
can be roughly $60MeV$ and a similar energy can be ascribed to the
corresponding motion.

\end{document}